# A Method for Accurate and Efficient Propagation of Satellite Orbits: A Case Study for a Molniya Orbit

Roberto Flores[a,b], Burhani Makame Burhani[a], Elena Fantino[a]*

[a] Department of Aerospace Engineering, Khalifa University of Science and Technology, Abu Dhabi, United Arab Emirates, P.O. Box 127788

[b] Centre Internacional de Mètodes Numèrics en Enginyeria (CIMNE), Gran Capità s/n, 08034 Barcelona, Spain

*Corresponding author: elena.fantino@ku.ac.ae

**Abstract**

Fast and precise propagation of satellite orbits is required for mission design, orbit determination and payload data analysis. We present a method to improve the computational performance of numerical propagators and simultaneously maintain the accuracy level required by any particular application. This is achieved by determining the positional accuracy needed and the corresponding acceptable error in acceleration on the basis of the mission requirements, removing those perturbation forces whose effect is negligible compared to the accuracy requirement, implementing an efficient and precise algorithm for the harmonic synthesis of the geopotential gradient (i.e., the gravitational acceleration) and adjusting the tolerance of the numerical propagator to achieve the prescribed accuracy level with minimum cost. In particular, to achieve the optimum balance between accuracy and computational performance, the number of geopotential spherical harmonics to retain is adjusted during the integration on the basis of the accuracy requirement. The contribution of high-order harmonics decays rapidly with altitude, so the minimum expansion degree meeting the target accuracy decreases with height. The optimum degree for each altitude is determined by making the truncation error of the harmonic synthesis equal to the admissible acceleration error. This paper presents a detailed description of the technique and test cases highlighting its accuracy and efficiency.

## 1 Introduction

Accurate predictions of satellite trajectories are required for mission analysis, trajectory design, targeting, guidance and navigation. The accelerated growth of the space industry, with missions of increased complexity, calls for better performance in such tasks as orbit propagation and determination. Furthermore, there is a critical need to predict the trajectories of space debris at the most populated altitudes [1]. The development of advanced tools to tackle effectively and accurately these problems is an inescapable requirement to maintain the current rate of progress.

There are three categories of orbit propagation methods: numerical, analytical, and semi-analytical [2, 3, 4]. Numerical methods, also termed special perturbations, compute and approximate solutions of the general equations of motion [see, e.g., 5, 6, 7, 8]. They are accurate and applicable to very complex scenarios, but computationally demanding. Analytical methods, also known as general perturbations, replace the original equations with an analytical approximation that captures the essential features of the problem. Fundamental analytical theories are due to [9, 10, 11, 12]. The approximate equations enable analytical integration, with substantial improvements in computational cost. The drawback is reduced accuracy. In a middle ground, semi-analytical methods blend numerical and analytical approaches by treating separately the short- and long-period components of the perturbations. The mean element rates are numerically propagated, whereas the short-period terms are modelled analytically by Fourier series. Among the most prominent semi-analytical theories we recall those due to [13, 14, 15].

The phenomenal increase in computer power over the last decades favored the investigation in the area of special perturbations, reducing the popularity of analytical theories, with the notable exception of SGP4 [16] which is widely used for tracking satellites and space debris in geocentric orbit. However, analytical techniques provide deeper insight into physical mechanisms at play. They are also essential tools for mission planning. More recently, the requirements introduced by the space law have resulted in applications with relaxed accuracy requirements (e.g., the long-term propagation of end-of-life disposal strategies) suitable for analytical and semi-analytical propagators. This is the context in which STELA (Semi-analytic Tool for End of Life Analysis) [17] and HEOSAT (Highly Elliptical Orbit SATellite propagator) [18] have been developed.

References [19, 20] review different types of Runge-Kutta integrators used for orbit propagation. Discussions on symplectic integration methods applied to celestial mechanics and astrodynamics can be found in [21, 22, 23, 24]. Formulations tailored for specific problems include the DROMO regularized propagator [25] and its developments [26, 27, 28]. [29] describes a fast numerical integration technique using a low-fidelity force model for tracking purposes.

The focus of the present contribution is simultaneously on the accuracy and the efficiency of orbital propagation. As discussed in [30], the numerical integration error is an important part of the overall uncertainty, and a tradeoff between computation time and accuracy must be established. In the same context, [31] investigated the sensitivity of the orbital dynamics with respect to geopotential truncation in a study devoted to the design of frozen orbits. Reference [32] analyzes the impact of geopotential expansion degree in the accuracy of orbit propagation for the LAGEOS (Laser Geodynamics Satellite) spacecraft. It also explores the effect of different choices of EGM (Earth Gravitational Model). We present a technique for the accurate and efficient propagation of geocentric orbits using Cowell's method [33]. The accuracy demanded by any particular application – derived from the mission requirements – is preserved by including in the equations of motion only the perturbing accelerations that produce deviations larger than the target accuracy. Also, the settings of the numerical integrator must be adjusted to keep discretization errors below the acceptable threshold while minimizing computational cost. We devote special care to the treatment of the acceleration caused by the geopotential, its expansion in spherical harmonics [34] and the accuracy of the Stokes coefficients of the EGM. The minimum expansion degree required to compute the gravitational acceleration is determined based on the desired accuracy level. The degree is adjusted dynamically as the calculation progresses, taking into account the instantaneous altitude of the satellite. The relation between the truncation error and the expansion degree is established beforehand, sampling the gravitational field at different latitudes, longitudes and heights. From this dataset, the required expansion degree for each altitude is determined as a function of the truncation error. Also, a statistical analysis of the gravitational field gives the maximum accuracy achievable at each height (i.e., the intrinsic uncertainty of the model when all the harmonics of the field are included). This information is valuable, as it determines whether or not a particular model of the gravity field (e.g., EGM96 [35], EGM2008 [36]) is satisfactory for a specific application.

This paper describes in detail a methodology originally developed to investigate station-keeping requirements of Tundra-type constellations [37]. We provide a comprehensive explanation of the steps required to adjust the physical model and numerical integrator in order to guarantee the accuracy of the solution, which are applicable to any type of orbit. We also compare the computational performance of the technique against the standard approach, which uses a fixed expansion degree for the gravitational field synthesis. This paper is an improved and extended version of a summary description presented at the International Astronautical Congress [38].

Section 2 of the paper describes the theoretical models and numerical algorithms used to facilitate reproduction of our results. Section 3 illustrates the analysis of the geopotential model uncertainties. Section 4 presents the method to determine, based on the accuracy requirements, the threshold acceleration level below which perturbations can be neglected. The practical application of the methodology is illustrated with a Molniya orbit [39] (where the large eccentricity highlights clearly the advantages of the dynamic expansion). Section 5 contains the numerical results. It gives guidelines for the correct setup of the integrator and presents the solution for the example orbit, focusing on the accuracy and computational performance. We include a comparison with simpler approaches as well as high-precision reference solutions to demonstrate the performance and accuracy of our technique. Finally, we draw the main conclusions in Section 6.

## 2   Material and methods

Our orbit propagation suite integrates the equations of motion in Cartesian coordinates with adaptive embedded Runge-Kutta (RK) schemes. For the calculations presented in this paper, we selected a scheme of orders 7 and 8 due to Fehlberg [40]. We used linear extrapolation, retaining the $8^{th}$ order solution.

The satellite state vector ($s_{sat}$) is formulated in the J2000 (noon January 1st 2000 Terrestrial Time) Mean Equator and Ascending Node reference frame. To a very high accuracy (23 milliarcseconds [41]) the orientation of this system coincides with the International Celestial Reference Frame, so it will be designated ICRF henceforth. The governing Ordinary Differential Equation (ODE) is

$$\frac{d\mathbf{s}_{sat}}{dt} = \frac{d}{dt}\begin{bmatrix}\mathbf{x}_{sat}\\ \mathbf{v}_{sat}\end{bmatrix} = \begin{bmatrix}\mathbf{v}_{sat}\\ \mathbf{a}_{sat}\end{bmatrix}, \qquad (1)$$

Where $\mathbf{s}_{sat}$ is the satellite state vector, containing the spacecraft position ($\mathbf{x}_{sat}$) and velocity ($\mathbf{v}_{sat}$). The satellite acceleration ($\mathbf{a}_{sat}$) can be split into a main component due to Earth's gravitational field ($\mathbf{g}$) and perturbative accelerations ($\mathbf{a}_{pert}$):

$$\mathbf{a}_{sat} = \mathbf{g} + \mathbf{a}_{pert}. \qquad (2)$$

The perturbations we shall consider in the analysis are: third-body gravitational perturbations of the Moon, Sun and Jupiter ($\mathbf{a}_{TB}$), solar radiation pressure ($\mathbf{a}_{rad}$), Earth albedo radiation pressure ($\mathbf{a}_{albe}$), solid Earth tides ($\mathbf{a}_{ET}$), ocean tides ($\mathbf{a}_{OT}$) and low-order relativistic correction ($\mathbf{a}_{rel}$):

$$\mathbf{a}_{pert} = \mathbf{a}_{TB} + \mathbf{a}_{rad} + \mathbf{a}_{albe} + \mathbf{a}_{ET} + \mathbf{a}_{OT} + \mathbf{a}_{rel}. \qquad (3)$$

We have not included aerodynamic drag in (3) because it is not relevant for the type of orbit we shall discuss. Nevertheless, its magnitude will be assessed in section 4.3 to confirm that it can be safely neglected. On the other hand, for Low Earth Orbit (LEO) applications atmospheric drag is extremely important and must be included in the analysis.

For the sake of clarity, and to limit the length of the paper, we have restricted the range of perturbations considered. However, the methodology we present is general and remains applicable to cases where additional forces play an important role. The procedure does not depend on the physical source of perturbations.

## 2.1 Third-body perturbations

The gravitational perturbations due to third bodies are given by their respective tidal forces. These are the sum of the gravitational pull on the satellite and the inertial force due to the acceleration of the origin of the reference frame (Earth's barycenter):

$$\mathbf{a}_{TB} = Gm_{TB}\left(\frac{\mathbf{x}_{TB} - \mathbf{x}_{sat}}{\|\mathbf{x}_{TB} - \mathbf{x}_{sat}\|^3} - \frac{\mathbf{x}_{TB}}{\|\mathbf{x}_{TB}\|^3}\right) \quad \text{where} \quad TB = Sun, Moon, Jupiter,\dots . \tag{4}$$

In Eq. (4), $\mathbf{x}_{TB}$ and $\mathbf{x}_{sat}$ denote the geocentric position vectors of the perturbing body and the satellite, respectively, while $m_{TB}$ is the mass of the third body and $G$ is the universal gravitational constant. For this work, the Sun, Moon and Jupiter have been included in the calculation. When the third body is very far from Earth – especially problematic for Jupiter, where $\|\mathbf{x}_{Jupiter}\| \gg \|\mathbf{x}_{sat}\|$ – expression (4) becomes ill-conditioned because it is the difference of two very similar vectors. This increases the relative error of the result compared with the uncertainty of each of the terms involved. This problem can be mitigated by reworking the formula of tidal acceleration as shown in [42]. The more robust expression – using the Sun as example – is:

$$\mathbf{a}_{Sun} = \frac{-Gm_{Sun}}{\|\mathbf{x}_{Sun}\|^3 (q+1)^{3/2}}\left(\mathbf{x}_{sat} + f\,\mathbf{x}_{Sun}\right), \tag{5}$$

where

$$q = \frac{\|\mathbf{x}_{sat}\|^2 - 2\,\mathbf{x}_{sat}\cdot\mathbf{x}_{Sun}}{\|\mathbf{x}_{Sun}\|^2} \quad;\quad f = q\,\frac{3 + 3q + q^2}{1 + (q+1)^{3/2}}. \tag{6}$$

To compute the position of the celestial bodies we used tabulated geocentric state vectors from JPL's Solar System Dynamics website [43]. To recover the position at an arbitrary date, the table is interpolated using cubic splines.

## 2.2 Computation of Earth's gravity field
### 2.2.1 Orientation of the terrestrial reference frame

To calculate the gravitational acceleration acting on the satellite its coordinates must be transformed to an Earth-fixed reference frame. Let us denote this reference as TRS (short for Terrestrial Reference System). This is a coordinate system with the *z* axis along the instantaneous axis of rotation of the Earth and the *x* axis pointing towards the reference meridian (Greenwich). The transformation from ICRF to TRS is a two-stage process that uses the Mean Equator and Ascending Node of Date reference frame –

MOD hereafter – as an intermediate step. The *z* direction of MOD also coincides with Earth's axis, while the *x* axis points towards the current vernal equinox (this frame is quasi-inertial).

For the transformation from ICRF to MOD we neglected nutation and polar motion. This introduces a misalignment smaller than 10" – 50 μrad – between the instantaneous axis of rotation of the Earth and the z axis of the MOD reference frame [41]. This error is deemed acceptable for our purposes. Therefore, we use only the Earth precession matrix $\mathbf{P}$ and the transformation is given by $\mathbf{x}^{MOD} = \mathbf{P} \cdot \mathbf{x}^{ICRF}$ [41]. Matrix $\mathbf{P}$ is expressed as a function of the Lieske et al. precession angles [44], computed with the polynomial approximation recommended by IAU (International Astronomical Union) [45].

Next, the transformation from MOD to TRS is a simple rotation about the *z* axis. The angle between MOD and TRS is the Greenwich Mean Sidereal Time (*GMST*). It can be expressed as

$$GMST = \theta(UT1) + \tau(TT) \,, \tag{7}$$

In the equation above $\theta$ represents the Earth rotation angle, which is a linear function of *UT1* – Universal Time, related to the apparent motion of the Sun – and $\tau$ accounts for precession of the equinoxes. *TT* – Terrestrial Time – has a direct relationship to physical time measured in SI seconds. UT1, however, cannot be accurately predicted because the rotation rate of the Earth is subject to very complex short-term variations. The difference between both timekeeping standards

$$\Delta T = TT - UT1 \tag{8}$$

is measured by the IERS (International Earth Rotation and Reference Systems Service). For the purpose of this paper (i.e., epochs less than 10 years in the future) a reasonable assumption is taking a constant rate of drift of *TT* with respect to *UT1*, equal to the average value over the last century. Using the tables in [46], we obtain

$$\frac{d \Delta T}{d TT} \approx 58 \frac{s}{century} \,, \tag{9}$$

which allows us to solve for GMST as a function of TT only. We used a polynomial approximation of relation (7) taken from [47], which is accurate to better than 1 mas (5 nrad). This error is negligible compared with the loss of accuracy incurred by removing nutation from the calculations.

### 2.2.2 Harmonic synthesis of the gravitational field

The gravitational acceleration (**g**) is given by the gradient of the geopotential (*V*)

$$\mathbf{g} = \nabla V = \frac{\partial V}{\partial x}\mathbf{i} + \frac{\partial V}{\partial y}\mathbf{j} + \frac{\partial V}{\partial z}\mathbf{k} = \frac{\partial V}{\partial r}\mathbf{e}_r + \frac{\partial V}{\partial \theta}\frac{\mathbf{e}_\theta}{r} + \frac{\partial V}{\partial \lambda}\frac{\mathbf{e}_\lambda}{r\sin\theta} \;, \tag{10}$$

depending on whether Cartesian $\{x, y, z\}$ or spherical $\{r, \theta, \lambda\}$ – radius, colatitude and longitude – coordinates are used. The potential is expressed as the sum of spherical harmonics [48]:

$$V(r,\theta,\lambda) = \frac{Gm_{Earth}}{r}\left(1 + \sum_{n=2}^{N}\left(\frac{a_{Earth}}{r}\right)^n \Omega_n\right), \tag{11}$$

where $N$ denotes the degree of the expansion, $a_{Earth}$ is Earth's equatorial radius and the partial sum of degree $n$ is given by

$$\Omega_n = \sum_{m=0}^{n} \bar{P}_{nm}(\theta)\left(\bar{C}^1_{nm}\cos m\lambda + \bar{C}^2_{nm}\sin m\lambda\right). \tag{12}$$

In Eq. (12) $\bar{P}_{nm}(\theta)$ is the fully-normalized Associated Legendre Function (ALF) of the first kind of degree $n$ and order $m$, and $\{\bar{C}^1_{nm}, \bar{C}^2_{nm}\}$ are the respective Stokes coefficients of the model. Expression (11) assumes the origin of the coordinate system coincides with Earth's barycenter, making the terms of degree 1 null.

Our propagator includes two different algorithms for computing the gradient (10). One, in spherical coordinates, implements the modified forward row recursion scheme from [48]. It is suitable for ultra-high-degree expansions ($N > 2000$). Through sequential memory access patterns and vectorization with AVX2 (Advanced Vector Extensions 2) instructions [49], the algorithm becomes very efficient in current CPU architectures. As an example, for $N = 30$ in single-threaded mode (i.e., utilizing only one core) an Intel i7-8750H mobile CPU – 2.2 GHz base clock, 3.9 GHz maximum boost – performs 730 000 field evaluations per second. While the forward row scheme is accurate and fast, it is not suitable for polar orbits because – in spherical coordinates – the derivative of the ALFs along the meridians is singular at the poles. To deal with completely arbitrary orbits, the propagator includes also a harmonic synthesis module that operates on Cartesian coordinates and is free from singularities. The algorithm [50] replaces the ALFs with Helmholtz polynomials $H_n^m$ [51, 52]. The field (11) is rewritten in terms of the direction cosines of the position vector

$$\chi_1 = \frac{x}{r} \;;\; \chi_2 = \frac{y}{r} \;;\; \chi_3 = \frac{z}{r} \;, \tag{13}$$

and the radial distance *r*. This yields a new series expansion of the form

$$V(\chi_1, \chi_2, \chi_2, r) = \sum_{n=0}^{N} p_n \sum_{m=0}^{n} D_{nm}(\chi_1, \chi_2) H_n^m(\chi_3) \ . \quad (14)$$

In the expression above, $p_n$ is the parallactic factor

$$p_n = \frac{Gm_{Earth}}{r} \left( \frac{a_{Earth}}{r} \right)^n , \quad (15)$$

and $D_{nm}$ is the mass coefficient function of degree *n* and order *m*

$$D_{nm}(\chi_1, \chi_2) = \overline{C}_{nm}^1 \, \mathrm{Re}\left( (\chi_1 + i\chi_2)^m \right) - \overline{C}_{nm}^2 \, \mathrm{Im}\left( (\chi_1 + i\chi_2)^m \right) \ . \quad (16)$$

Details on the algorithm as well as efficient recursion schemes for computing the sums can be found in [53].

### 2.2.3 Secular evolution of the gravitational field

For the modelling of Earth's gravitational field we followed the guidelines from the IRS Conventions 2010 standard (IERS Technical Note No. 36, IERS TN36 henceforth) [54]. It establishes EGM2008 as the recommended geopotential model with some adjustments.

To account for the progressive drift of the low-degree zonal harmonics, IERS TN36 gives their values at a reference epoch ($\overline{C}_{n0}^{t_0}$) and a secular rate ($\mathrm{d}\overline{C}_{n0}/\mathrm{d}t$):

$$\overline{C}_{n0}^1 = \overline{C}_{n0}^{t_0} + \frac{\mathrm{d}\overline{C}_{n0}}{\mathrm{d}t}(t - t_0) \qquad n = 1, 2, 3 \ , \quad (17)$$

where $t_0$ denotes the reference epoch (J2000). Furthermore, the IERS TN36 recommended value of $\overline{C}_{20}^{t_0}$ differs from the original EGM2008 datum. It is based on 17 years of SLR (Satellite Laser Ranging) which is expected to improve on the 4 years of GRACE (Gravity Recovery and Climate Experiment) data used in EGM2008.

### 2.3 Solid Earth tide modeling

Due to the combined tidal forces of the Moon and Sun, the figure of the Earth experiences cyclical variations that cause a modulation of the gravitational field. This effect has been accounted for using the procedure defined in IERS TN36. The change in the gravitational field is expressed as a variation of the Stokes coefficients given by

$$\Delta \bar{C}_{nm}^1 - i\Delta \bar{C}_{nm}^2 = \frac{k_{nm}}{2n+1} \sum_{j=Moon,Sun} \frac{m_j}{m_{Earth}} \left(\frac{a_{Earth}}{r_j}\right) \bar{P}_{nm}(\theta_j) e^{-im\lambda_j} \quad n = 2,3 \ ; \ m = 0,1,..,n \ , \quad (18)$$

where $i$ denotes the imaginary unit, $j$ is either the Sun or Moon and $k_{mn}$ is the nominal Love number for degree $n$ and order $m$. $\{r_j, \theta_j, \lambda_j\}$ represent the geocentric spherical coordinates of celestial body $j$ in the TRS frame and $m_j$ is its mass. We used the Love numbers corresponding to the anelastic Earth model, which includes the phase lag between the gravity excitation and Earth's deformation. There is an additional effect caused by the degree 2 tides on the degree 4 Stokes coefficients:

$$\Delta \bar{C}_{4m}^1 - i\Delta \bar{C}_{4m}^2 = \frac{k_{2m}^+}{5} \sum_{j=Moon,Sun} \frac{m_j}{m_{Earth}} \left(\frac{a_{Earth}}{r_j}\right) \bar{P}_{nm}(\theta_j) e^{-im\lambda_j} \quad m = 0,1,2 \ . \quad (19)$$

In a second step, the coefficient variations computed from (18) and (19), which assume fixed Love numbers, are corrected for frequency dependency. For more details, refer to IERS TR36. Finally, because we used the zero-tide version of EGM2008, the time-average of $\Delta \bar{C}_{20}^1$ must be subtracted from the variation, because it is already included in the geopotential model [54].

## 2.4 Ocean tide modeling

The tidal displacement of the oceans also influences Earth's gravity field. We modeled this effect with the FES2004 (Finite Element Solution 2004) ocean tide model [55]. The variation of the Stokes coefficients is obtained as a sum of the contributions of the different tide constituents:

$$\Delta \bar{C}_{nm}^1 - i\Delta \bar{C}_{nm}^2 = \sum_f \sum_+ \left(\bar{C}_{f,nm}^{1\pm} \mp i \bar{C}_{f,nm}^{2\pm}\right) e^{\pm i\theta_f(t)} \ , \quad (20)$$

where $f$ denotes a constituent and $\bar{C}_{f,nm}^{1\pm}$ and $\bar{C}_{f,nm}^{2\pm}$ are the amplitudes of the corresponding geopotential harmonics of degree $n$ and order $m$. The superscripts + and – indicate the prograde and retrograde components of the main wave $f$ and $\theta_f(t)$ is the argument of the tide constituent. More details of the theory behind Eq. (20) can be found in [54].

For the scope of this paper, including the main components of the solid Earth and ocean tides is sufficient. Higher-accuracy applications, especially in LEO, might require considering solid Earth pole tides and ocean pole tides as described in IERS TR36.

## 2.5 Solar radiation pressure perturbation

The effect of radiation pressure is included with a simplified model that treats the satellite as a 100% reflective sphere. The net acceleration is

$$\mathbf{a}_{rad} = -\phi \frac{W_{Sun}}{4\pi c} \frac{A_{sat}}{m_{sat}} \frac{\mathbf{x}_{sS}}{\|\mathbf{x}_{sS}\|^3} \ , \tag{21}$$

where $W_{Sun}$ is the absolute luminosity of the Sun – $3.83 \cdot 10^{26}$ W – $c$ denotes the speed of light in vacuum, $A_{sat}/m_{sat}$ is the area-to-mass ratio of the satellite and $\mathbf{x}_{sS} = \mathbf{x}_{Sun} - \mathbf{x}_{sat}$. The eclipse factor $\phi$ accounts for occultation of the Sun by the Earth. To determine its value, the angular radii of the Earth and Sun viewed from the satellite must be determined:

$$\alpha_{Earth} = \arcsin\left(\frac{a_{Earth}}{\|\mathbf{x}_{sE}\|}\right) \ ; \ \alpha_{Sun} = \arcsin\left(\frac{a_{Sun}}{\|\mathbf{x}_{sS}\|}\right) \approx \frac{a_{Sun}}{\|\mathbf{x}_{sS}\|} \ , \tag{22}$$

where $a_{Sun}$ is the Solar radius – $6.97 \cdot 10^5$ km – and $\mathbf{x}_{sE} = \mathbf{x}_{Earth} - \mathbf{x}_{sat}$. In (22) we assumed $a_{Sun} \ll \|\mathbf{x}_{sS}\|$, which is reasonable for a spacecraft orbiting Earth. Next, the angular separation $\beta$ between Sun and Earth is computed:

$$\beta = \frac{\arccos(\mathbf{x}_{sE} \cdot \mathbf{x}_{sS})}{\|\mathbf{x}_{sE}\| \|\mathbf{x}_{sS}\|} \ , \tag{23}$$

which discriminates between three different configurations

$$\begin{aligned} \beta > (\alpha_{Earth} + \alpha_{Sun}) &\rightarrow \phi = 1 &\text{(no eclipse)} \\ (\alpha_{Earth} - \alpha_{Sun}) < \beta < (\alpha_{Earth} + \alpha_{Sun}) &\rightarrow \phi \in \,]0,1[ &\text{(satellite in penumbra)} \\ \beta < (\alpha_{Earth} - \alpha_{Sun}) &\rightarrow \phi = 0 &\text{(satellite in umbra)} \end{aligned} \tag{24}$$

In case of partial eclipse – penumbra – the exact expression for the fraction of the solar disk shadowed by the Earth can be found in [56]. It is, however, cumbersome and costly to evaluate. For most Earth orbits we can safely assume $\alpha_{Earth} \gg \alpha_{Sun}$, in which case it is easy to show that

$$\phi \approx 1 + \frac{\varphi\sqrt{1-\varphi^2} - \arccos\varphi}{\pi}, \tag{25}$$

with $\varphi = (\beta - \alpha_{Earth})/\alpha_{Sun}$.

## 2.6 Albedo radiation pressure perturbation

For low and moderate height orbits the radiation pressure from light reflected by the Earth surface can also play a relevant role. We used a model for a spherical satellite that has the advantage of reducing to simple algebraic expressions for those cases where the terminator (the boundary between the shadowed and the sunlit parts of Earth) is not visible from the spacecraft [57]. When the terminator is in sight, the acceleration can be computed using line integrals instead of surface integrals, reducing the computational burden. The model assumes a uniform albedo across Earth's surface and that each illuminated point behaves like a Lambertian source. Thus, the radiance ($L$) of each illuminated surface element is

$$L = \rho_{Earth} \frac{W_{Sun}}{4\pi \|\mathbf{x}_{Sun}\|^2} \frac{\cos\Psi_{Sun}}{\pi}, \tag{26}$$

where $\rho_{Earth}$ denotes Earth's average albedo (we used a value of 0.3), the second term of the product in the RHS is the incident solar flux at Earth's surface and $\Psi_{Sun}$ is the incidence angle of the Sun rays. The acceleration experienced by the spacecraft due to the radiation reflected from a surface element of the Earth is:

$$d\mathbf{a}_{albe} = \frac{A_{sat}}{m_{sat}} \frac{L\,d\Omega}{c} \mathbf{u}. \tag{27}$$

In Eq. (27) $d\Omega$ represents the solid angle subtended by the surface element seen from the satellite and $\mathbf{u}$ is a unit vector pointing from the surface element to the satellite. Integrating Eq. (27) over the visible illuminated part of Earth's surface gives the perturbing acceleration. For details of the integration process, refer to [57].

Given the scope of this paper, the simple model of radiation pressure outlined in sections 2.4 and 2.6 is adequate. However, depending on the application, additional effects might be considered such as: infrared radiation from Earth's surface, the satellite's own thermal radiation distribution and the thrust due to transmitter antennas [58][59].

## 2.7 Relativistic correction

To the lowest order approximation, the relativistic correction is an additional attraction given by [60]

$$\mathbf{a}_{rel} = -\frac{3Gm_{Earth}h_{sat}^2}{c^2}\frac{\mathbf{x}_{sat}}{\|\mathbf{x}_{sat}\|^5} \ , \qquad (28)$$

where $h_{sat} = \|\mathbf{x}_{sat} \times \mathbf{v}_{sat}\|$ is the magnitude of the specific orbital angular momentum of the satellite. Expression (28) is based on the Schwarzschild central field only, it does not include geodetic precession (de Sitter effect) or gravitomagnetism (frame dragging). For the purposes of this paper the low-order approximation is sufficient, but a more detailed formulation may be required in higher-accuracy applications. For a comprehensive treatment, please refer to IERS TN36.

## 3  Intrinsic geopotential model uncertainty and truncation error

When propagating Earth orbits in a non-spherical gravitational field, especial attention must be paid to the choice of geopotential model and expansion degree. Depending on the accuracy requirements of each particular application, the admissible error when evaluating the gravity acceleration changes; there is no "one size fits all" solution. Unfortunately, this is often ignored in existing literature, with the expansion degree of the harmonic series chosen beforehand, without giving supporting evidence. To guarantee the accuracy of a propagated trajectory, three steps are required to choose a suitable geopotential model:

1. Determine the accuracy in the evaluation of the spacecraft acceleration suitable for a particular application

2. Select an EGM – e.g., EGM96, EGM2008 or EIGEN-6C4 (European Improved Gravity model of the Earth by New techniques [61]) – adequate for the accuracy determined in step 1

3. Determine the minimum expansion degree required in order to keep the errors within tolerance but without incurring in excessive computational cost.

Section 4 of the paper will deal with step 1, while this part focuses on items 2 and 3. For the time being, we shall assume that the absolute accuracy ($\varepsilon_a$) required for the computation of the acceleration is known.

### 3.1 Intrinsic accuracy of the geopotential model

With "intrinsic" accuracy we mean the best precision that can be obtained from a particular geopotential model. This is the uncertainty of the acceleration when all the available Stokes coefficients are included in the expansion (i.e., the series is not truncated). Given that the coefficients are measured experimentally, they have an intrinsic uncertainty characterized by their standard deviation $\sigma_{nm}^\alpha = \sigma\left(\overline{C}_{nm}^\alpha\right)$. The matrix $\sigma_{nm}^\alpha$ is part of the definition of a geopotential model and is publicly available. The potential expansion (14) can be recast as

$$V(x,y,z) = \sum_{\forall n,m,\alpha} \overline{C}_{nm}^\alpha A_{nm}^\alpha(x,y,z) \quad \text{where} \quad \alpha = 1,2. \tag{29}$$

We used Cartesian coordinates in (29) to make the expressions for the acceleration simpler, but our discussion would be equally valid for the expansion in spherical coordinates (11). For any point in space, the coefficients $\overline{C}_{nm}^\alpha$ on the Right Hand Side (RHS) of (29) have an associated uncertainty $\sigma_{nm}^\alpha$. On the other hand, the terms $A_{nm}^\alpha(x,y,z)$ – combinations of ALFs, trigonometric and arithmetic functions – can, in principle, be evaluated exactly. Applying the Central Limit Theorem [62] to Eq. (29), assuming that $\overline{C}_{nm}^\alpha$ are randomly-distributed independent variables while the functions $A_{nm}^\alpha$ are exact, gives

$$\sigma(V) = \sqrt{\sum_{\forall n,m,\alpha} \left(\sigma_{nm}^\alpha A_{nm}^\alpha\right)^2} = \sqrt{\sum_{\forall n,m,\alpha} S_{nm\alpha}}, \tag{30}$$

where $S_{nm\alpha}$ denotes the variance of the $\overline{C}_{nm}^\alpha A_{nm}^\alpha$ terms. Thus, it is possible to compute the uncertainty of the potential at each point in space. For trajectory propagation, we are more interested in the uncertainty of the gravitational acceleration. Proceeding in the same way as for the potential:

$$s(g_i) = \sum_{\forall n,m,\alpha} S_{nm\alpha}^i \quad i = 1,2,3, \tag{31}$$

where $s(g_i)$ is the variance of the i-th component of the gravitational acceleration and

$$S^i_{nm\alpha} = \left( \sigma^\alpha_{nm} \frac{\partial A^\alpha_{nm}}{\partial x_i} \right)^2 . \qquad (32)$$

The standard deviation of the acceleration vector can be estimated, using the Euclidean norm, as

$$\sigma_g = \sqrt{s(g_1) + s(g_2) + s(g_3)} . \qquad (33)$$

Using a 95% confidence level – $2\sigma$ – we can consider a particular model adequate if

$$2\sigma_g < \varepsilon_a . \qquad (34)$$

The reader must keep in mind that $\sigma_g$ is a function of position, because the terms $A^\alpha_{nm}$ vary in space. Therefore, condition (34) is a local requirement. It is impractical to test (34) for every point along the trajectory, hence a condition that can be checked before starting the propagation is needed. To this end, one must realize that the largest changes in $\sigma_g$ are due to altitude, with longitude and latitude being of secondary importance. High-degree harmonics decay rapidly with height. Therefore, far from the surface the uncertainty is controlled by the low-degree terms of the expansion. To determine the typical uncertainty at each altitude, we can sample the function $\sigma_g$ (33) with a sufficiently fine grid in latitude and longitude, retaining the maximum value to obtain a conservative estimate. An exact value is not needed because, to estimate errors, what really matters is the order of magnitude of the uncertainty. Let us consider a grid of points in spherical coordinates

$$\mathbf{x}_{ijk} : \{h_i, \theta_j, \lambda_k\} \quad i = 1,...,n_h \; ; \; j = 1,...,n_\theta \; ; \; k = 1,...,n_\lambda \quad , \qquad (35)$$

where $h = r - a_{Earth}$ is the geocentric altitude. The maximum value of $\sigma_g$ among all the sampling points at a given altitude is used to derive the uncertainty estimate

$$\tilde{\varepsilon}^{\min}_g(h_i) = 2\max\left[\sigma_g(\mathbf{x}_{ijk})\right] \quad \forall \, j,k . \qquad (36)$$

We use the superscript min because the quantity denotes the minimum uncertainty achievable if all the terms of the field are included. The factor 2 on the RHS of (36) comes from the 95% confidence level criterion (34). If the series is truncated, the uncertainty may be larger. $\tilde{\varepsilon}^{\min}_g(h)$ can be precomputed and stored in a table. Then, to determine if a geopotential model is adequate, we only need to replace $2\sigma_g$ in (34) with the conservative estimate (36), giving

$$\tilde{\varepsilon}_g^{\min}(h) < \varepsilon_a \quad \forall \, h \in \left[h_{per}, h_{apo}\right], \tag{37}$$

where the subscripts *per* and *apo* denote the perigee and apogee of the orbit. We calculated $\tilde{\varepsilon}_g^{\min}$ for EGM96 and EGM2008, using the zero-tide version of the models. EGM96 is now deprecated, but was the recommended standard before ERS TN36 was released. The difference between the two models is illustrative of the progress of technology. The results are shown in Fig. 1 for a uniform grid in spherical coordinates with 10º resolution in both longitude and latitude. Tests with coarser grids showed no significant differences, meaning the sample is representative. The plot indicates that, over the range of altitudes tested, EGM2008 yields a tenfold accuracy improvement over EGM96. At altitudes between 1000 and 2000 km the difference is even larger, with EGM2008 having an advantage of two orders of magnitude.

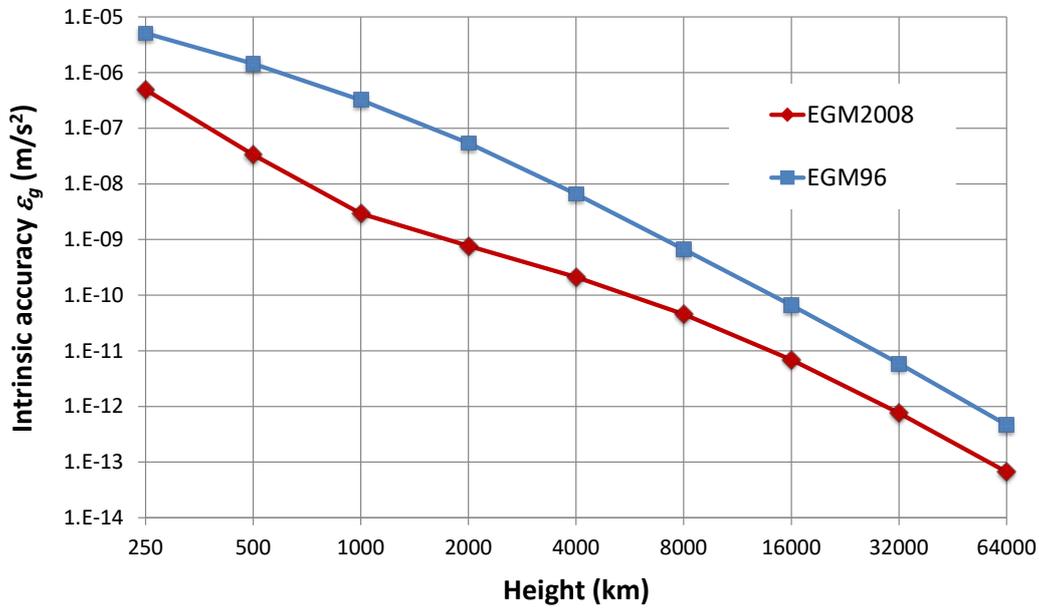

**Fig. 1 - Intrinsic accuracy vs. height for EGM96 & EGM2008 (10º grid)**

Over the range of altitudes tested, the accuracy of both model changes by 7 orders of magnitude. The minimum uncertainty corresponds to maximum height, where the value of the field is controlled by the low-degree harmonics, which can be measured with high accuracy. Closer to the surface, the role of high degree harmonics becomes important and, given their larger uncertainty, the error increases rapidly. Fig. 1 offers a simple way of determining the suitability of a geopotential model for a specific application. For example, imagine that an accuracy of $10^{-8}$ m/s$^2$ is sought at *h*=1000 km. The chart shows that EGM2008 is adequate but EGM96 is not. On the other hand, if the same accuracy is required for a geostationary satellite − *h* =35 786 km − either model is suitable.

## 3.2 Truncation error of the field expansion

The intrinsic accuracy $\tilde{\varepsilon}_g^{min}$, by itself, only determines if it is possible to reach the target accuracy with a given model. To achieve maximum computational efficiency while respecting tolerances, the truncation error of the field expansion must be characterized. Let $N_{max}$ be the maximum field expansion degree for a certain geopotential model (e.g., $N_{max}$=2159 for EGM2008). If the series (11) is truncated at degree $N<N_{max}$, we define the absolute truncation error of the gravitational acceleration as

$$\mathbf{\varepsilon}_{tru}(\mathbf{x}, N) = \mathbf{g}(\mathbf{x}, N) - \mathbf{g}(\mathbf{x}, N_{max}) \ . \tag{38}$$

Just like (33), expression (38) is not very useful in practice because it is a function of height, longitude and latitude. It is also very expensive to compute, as it requires expanding the field all the way to $N_{max}$, at which point the truncation error vanishes anyway. Thus, a fast way to estimate $\varepsilon_{tru}$ is needed. Remembering that the relative importance of the terms in the harmonic series is governed mainly by altitude, we can sample the truncation error at the points of a grid and compute an estimate of the typical value:

$$\tilde{\varepsilon}_{tru}^l(h_i, N) = \max\left[\left|\mathbf{\varepsilon}_{tru}(\mathbf{x}_{ijk}, N) \cdot \mathbf{e}^l\right|\right] \quad \forall j,k \quad l=1,2,3 \ . \tag{39}$$

In (39) $\tilde{\varepsilon}_{tru}^l$ denotes the *l-th* component of the typical error and $\mathbf{e}^l$ are the unit vectors of the reference frame. As an example, Fig. 2 shows the three components of the truncation error as a function of the expansion degree for an altitude of 2000 km. The intrinsic accuracy of the EGM2008 model is also included for reference. For expansion degrees above 44, the intrinsic uncertainty of the model becomes dominant.

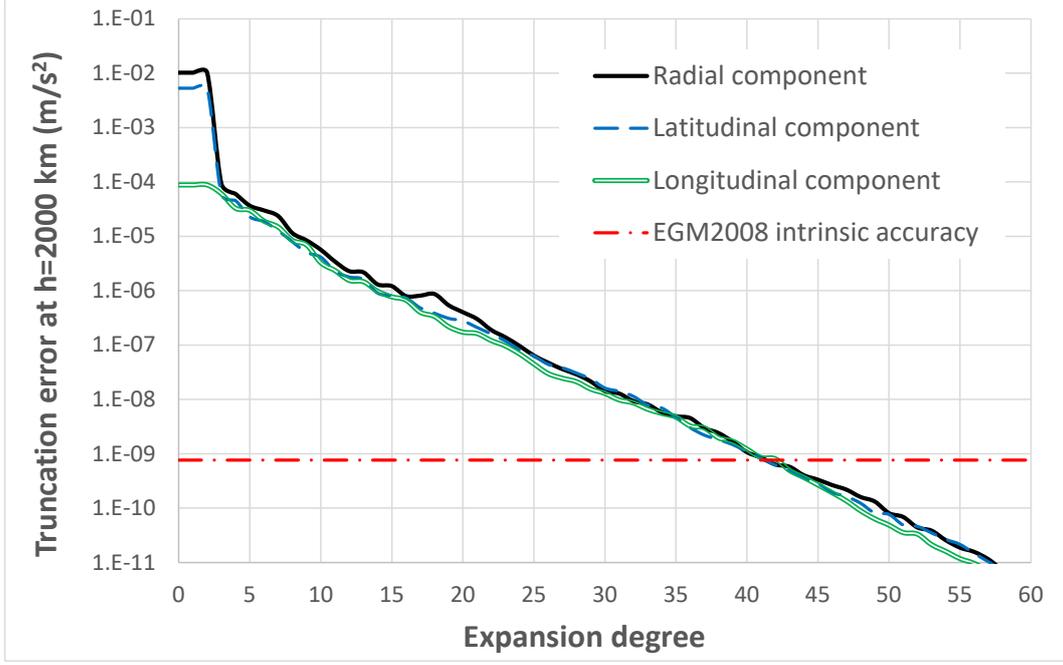

Fig. 2 – Truncation error of the three acceleration components at 2000 km altitude for the EGM2008 geopotential model

While the components of the truncation error are interesting for illustrative purposes, the magnitude of the error vector is more convenient for estimating propagation uncertainties. Therefore, moving forward, we shall retain only the norm of the truncation error:

$$\tilde{\varepsilon}_{tru}(h_i, N) = \max\left[\left\|\boldsymbol{\varepsilon}_{tru}(\mathbf{x}_{ijk}, N)\right\|\right] \quad \forall\, j, k \ . \tag{40}$$

From the computational point of view, the most useful relation is the reciprocal of (40), giving the minimum required expansion degree ($N_{req}$) as a function of the altitude and truncation error

$$N_{req}(h, \tilde{\varepsilon}_{tru}) \ . \tag{41}$$

This function is easy to obtain once (40) has been evaluated. Fig. 3 shows the variation of $N_{req}$ for the EGM2008 model using the same grid as Fig. 1. In a double-log plot, the $N_{req}$-$h$ curves for constant truncation error are almost linear, with small irregularities due to the discrete nature of the expansion degree, which can only take integer values. The slope of the curves indicates that, at constant accuracy, the expansion degree must be doubled each time the height is halved. Note that the graph shows $N_{req} = 1$ for $h = 64000$ km and $\varepsilon_{tru} = 10^{-5}$ m/s². As indicated in Sect. 2.2, due to the choice of the origin of coordinates, the Stokes coefficients of degree 1 are null. Therefore, $N_{req} = 1$ is equivalent to $N_{req} = 0$ (i.e., the gravity field of a point mass). $N_{req}=1$ has been retained because zero cannot be plotted in a logarithmic scale.

An interesting result can be obtained equating the intrinsic accuracy (36) to the truncation error(40):

$$\tilde{\varepsilon}_g^{\min}(h) = \tilde{\varepsilon}_{tru}(h, N_{eff}),  \qquad (42)$$

where $N_{eff}(h)$ is the maximum effective expansion degree. It represents the degree that makes the intrinsic accuracy and the truncation error comparable. Adding terms of degree higher than $N_{eff}$ would not improve the accuracy of the acceleration because the uncertainty of the Stokes coefficients would dominate the error.

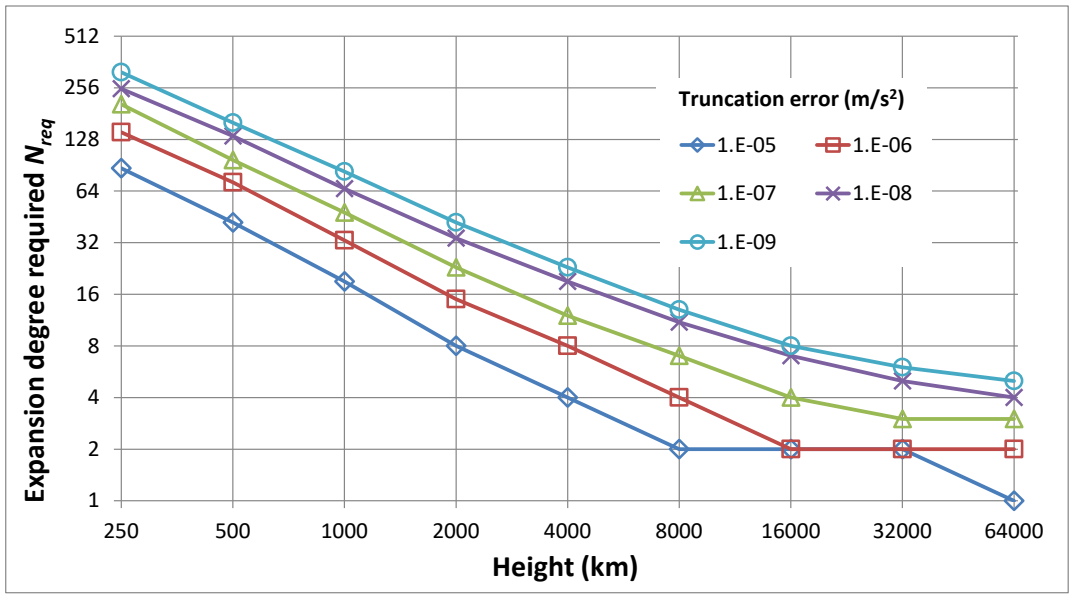

**Fig. 3 - Required expansion degree vs. height and truncation error for EGM2008 (10º grid)**

On the other hand, if the truncation error is large compared to the model uncertainty ($\tilde{\varepsilon}_g^{\min}$), the latter can be neglected. Using an engineering approach, we assume the truncation error dominates when it is 10 times larger than $\tilde{\varepsilon}_g^{\min}$. Denoting the corresponding degree with $N_{tru}$

$$\tilde{\varepsilon}_{tru}(h, N_{tru}) = 10\, \tilde{\varepsilon}_g^{\min}(h). \qquad (43)$$

The values of $N_{eff}$ and $N_{tru}$ as functions of altitude can be found in Fig. 4. Using an expansion degree higher than $N_{eff}$ is a waste of computational resources and should be avoided. From Fig. 4 we learn, for example, that in geostationary orbits it is pointless to use $N$ above 8, which corresponds to an accuracy of $10^{-12}$ m/s². The only way to reduce the uncertainty would be switching from EGM2008 to an improved geopotential model. However, whenever $N<N_{tru}$, the uncertainty of the Stokes coefficients can be safely ignored and the accuracy of the harmonic synthesis can be estimated directly from Fig. 3.

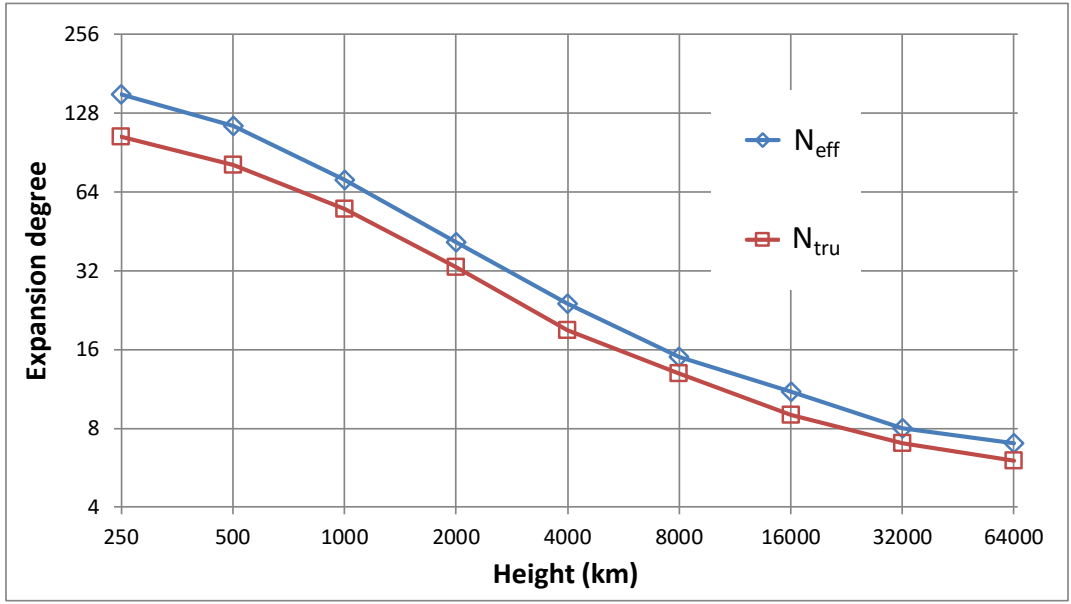

**Fig. 4 - Effective and truncation error-dominated expansion degrees (EGM2008)**

Given the steep rate of change of $N_{req}$ with height, it is clear that large computational savings can be obtained for eccentric orbits if instead of using a fixed expansion degree, $N$ is adjusted dynamically along the trajectory. This can be done inexpensively because $N_{req}(h,\varepsilon)$ only needs to be computed once and stored in a table. Then, during propagation, the table is interpolated to yield the optimum expansion degree at all times.

## 4 Determination of the accuracy requirements and numerical setup

In this section we present a systematic method for establishing the accuracy requirements, streamlining the physical model and enforcing the numerical tolerances while maximizing efficiency. We use a Molniya orbit to illustrate the methodology. Due to its large eccentricity, it covers a wide range of altitudes. This requires extra attention setting up the calculation, but allows for large performance gains if the model is tuned correctly.

## 4.1 A practical case: the Molniya orbit

Molniya orbits (see Fig. 5) were developed to provide communications and remote sensing services to high-latitude areas of the Soviet Union. These regions cannot be served effectively with geostationary satellites due to insufficient elevation over the horizon. A Molniya orbit has a period of one half sidereal day[1] ($T$=43082.05 s), critical inclination (63.4º) to reduce drift of the argument of perigee due to Earth's polar flattening, large eccentricity (~0.75) and argument of perigee of -90º (i.e., the apogee coincides with the point of highest latitude). Due to the high eccentricity, the satellite spends most of the time close to the apogee – this is called apogee dwell – which is placed above the zone to be served. Using a constellation of three satellites in evenly spaced orbital planes, it is possible to provide continuous coverage to the area of interest. The perturbation forces will cause the orbital elements of the satellite to drift with time, requiring periodic station-keeping maneuvers. The propellant budget for these maneuvers is a fundamental design constraint, as it limits the operational life of the satellite.

For our example, let us assume that the satellite operations require a 1º positional accuracy. Molniya satellites do not remain fixed in the sky and therefore do not need the same orbit stability as geostationary spacecraft (which are typically maintained inside a 0.1º window). An angular displacement of 1º corresponds to a shift in the ground track of some 100 km. This is a small distance compared with the size of the coverage area, typically spanning thousands of kilometers. We shall assume that the station-keeping maneuvers are scheduled once per month (once every 15 days is a common practice for geostationary satellites). Therefore, in order to plan the maneuvers, the operator must be able to sense a drift of 1º over a period of 30 days. We shall further assume that the propagation error must be at least two orders of magnitude smaller to predict reliably if a maneuver will be necessary and determine the amount of propellant required. The accuracy target for the propagator will then be set to 0.01º over 30 days. Using this condition as starting point, we shall fine-tune the model to achieve maximum performance without compromising the accuracy. To this effect, the initial orbital parameters of the satellite must be defined. We have chosen the following values:

---

[1] In reality, the orbital period differs slightly — by a few seconds — from one half sidereal day. This reduces the drift of the ground track when the ascending node of the orbit precesses due to Earth's oblateness. However, this detail is not relevant for our discussion. See [37] for more information.

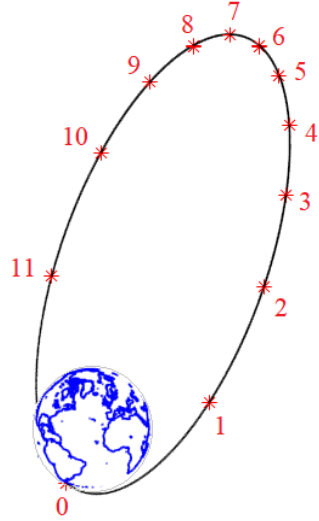

**Fig. 5 - Molniya orbit. The positions of the satellite at one-hour intervals are marked**

- Semimajor axis $a_0 = 26562.85$ km (determined from the orbital period)
- Eccentricity $e_0 = 0.7222$
- Inclination $i_0 = 63.4°$ (critical)
- Right Ascension of the Ascending Node (RAAN) $\Omega_0 = 0°$
- Argument of the perigee $\omega_0 = 270°$
- Initial true anomaly $\theta_0 = 180°$ (apogee)
- Start of propagation JD 2459215.5 (January 1st 2021, 00:00 UT)

The values of $\Omega_0$ and $\theta_0$ are arbitrary and do not play any important role for the objectives of the present discussion. They influence the position of the ground track, but do not have any effect on the physics or numerical complexity of the problem. From the orbital parameters, we can compute the perigee and apogee radii – $r_{per} = a_0(1-e_0) = 7379$ km, $r_{apo} = a_0(1+e_0) = 45747$ km – which play a very important role in the estimation of the magnitude of the different perturbations.

The area-to-mass ratio of the satellite is required to compute the acceleration due to radiation pressure. We shall assume $A_{sat}/m_{sat} = 0.01$ m²/kg, which is sensible for a communications satellite.

## 4.2 Determination of the acceleration threshold

A cumulative angular error of 0.01° over 30 days for a satellite with an orbital period of one half sidereal day is equivalent to a drift of 3 μrad per orbit. To make this equivalence, we implicitly assumed that the drift of the orbit is linear in time. As long as the perturbations are small (i.e., the orbit does not deviate significantly from its nominal configuration, something which is true in real operations) this is a reasonable approximation. It is the basis, for example, of Gauss' planetary equations [63], which express the rate of change of the orbital elements as linear functions of the perturbative force. While the rates change continuously along the orbit, their net effect integrated over one revolution is much more stable. Therefore, we expect the order of magnitude of the change per orbit to be consistent.

To convert from angular error to distance uncertainty, we shall assume the most disadvantageous case (i.e., satellite at the perigee). An apparent displacement of 3 μrad in the sky when the altitude of the satellite is $h_{per} = 1000\,\text{km}$ corresponds to a linear displacement of 3 m (or less). Therefore, if the positional accuracy is $\varepsilon_x = 3\,\text{m}$, the angular error will be smaller than the target – 3 μrad – over the complete orbit. This is a conservative approach because, as far as communications service is concerned, the important area is the apogee. The same angular uncertainty when the satellite is near maximum altitude corresponds to a much larger position error.

Once the positional accuracy requirement has been stablished, it is possible to determine the acceleration threshold $\varepsilon_a$. This is the maximum error in the determination of the perturbative accelerations that does not compromise the accuracy of the solution. Let $\mathbf{x}(t)$ be the – hypothetical – exact trajectory, computed with a perfect physical model, whereas $\tilde{\mathbf{x}}(t)$ represents the approximate solution obtained with simplified physics. The differential equations governing these two variables are:

$$\begin{aligned} \ddot{\mathbf{x}} &= \mathbf{a}(\mathbf{x},t) \\ \ddot{\tilde{\mathbf{x}}} &= \tilde{\mathbf{a}}(\tilde{\mathbf{x}},t) = \mathbf{a}(\tilde{\mathbf{x}},t) + \mathbf{a}_e(\tilde{\mathbf{x}},t) \end{aligned}, \qquad (44)$$

where $\mathbf{a}$ and $\tilde{\mathbf{a}}$ denote the exact and approximate acceleration fields and $\mathbf{a}_e$ is the error of the latter. Therefore, the differential equation for the positional error $\mathbf{e}(t) = \tilde{\mathbf{x}} - \mathbf{x}$ is

$$\ddot{\mathbf{e}} = \left[\mathbf{a}(\tilde{\mathbf{x}},t) - \mathbf{a}(\mathbf{x},t)\right] + \mathbf{a}_e(\tilde{\mathbf{x}},t) , \qquad (45)$$

with the initial conditions (we assume that initial position and velocity are exact)

$$t = 0 \rightarrow \begin{cases} \mathbf{e} = \mathbf{0} \\ \dot{\mathbf{e}} = \mathbf{0} \end{cases}. \qquad (46)$$

The attentive reader has probably realized that Eq. (45) is very similar to Encke's method of special perturbations [64] if the perturbative acceleration is replaced with the error term $\mathbf{a}_e$. The bracket in the RHS of Eq. (45) is a feedback term that couples with the basic error due to $\mathbf{a}_e$. Once the satellite moves away from the ideal trajectory, there is an additional change in acceleration due to the non-uniformity of the force field that contributes to the uncertainty. In principle, integrating twice equation (45) would yield the position error

$$\mathbf{e}(t) = \int_0^t \int_0^\eta \ddot{\mathbf{e}}(\tilde{\mathbf{x}}, \xi) \, d\xi \, d\eta \ . \tag{47}$$

In a real situation, however, expression (47) cannot be evaluated directly because the acceleration error $\mathbf{a}_e$ is unknown. Nevertheless, we can estimate the size of $\mathbf{e}$ by looking at the magnitude of the relevant terms. Let $\varepsilon_a \geq \|\mathbf{a}_e\|$ be an upper bound of the magnitude of the acceleration error. The simplest error estimate is obtained assuming $\|\mathbf{a}(\tilde{\mathbf{x}}, t) - \mathbf{a}(\mathbf{x}, t) + \mathbf{a}_e\| \sim \|\mathbf{a}_e\|$ (i.e., $\mathbf{a}_e$ is dominant or, at least, comparable to the feedback term). In that case:

$$\|\mathbf{e}\| \approx \left\| \iint \mathbf{a}_e \right\| \leq \iint \|\mathbf{a}_e\| \leq \iint \varepsilon_a = \varepsilon_a \frac{t^2}{2} \ . \tag{48}$$

Note that (48) yields a conservative bound of $\|\mathbf{e}\|$, because neither the magnitude nor the direction of $\mathbf{a}_e$ are constant in reality. Most orbital perturbations are cyclic in nature, so their net effect over one revolution is smaller. Setting $\|\mathbf{e}\| = \varepsilon_x$ (the target positional accuracy after one orbit) and $t = T$ in (48) gives an estimate of the acceleration threshold

$$\varepsilon_a \sim \frac{2\varepsilon_x}{T^2} = 3.3 \cdot 10^{-9} \ \frac{\mathrm{m}}{\mathrm{s}^2} \ . \tag{49}$$

Error estimations are never accurate, only their orders of magnitude are relevant. Therefore, we can set $\varepsilon_a = 10^{-8} \ \mathrm{m/s^2}$ (because $\log_{10} 3.3 > 0.5$). Rounding the value up also compensates for the conservative bias of the estimate.

Once the order of magnitude of the admissible error has been determined, it is good practice to check that the condition $\|\mathbf{a}(\tilde{\mathbf{x}}, t) - \mathbf{a}(\mathbf{x}, t) + \mathbf{a}_e\| \sim \|\mathbf{a}_e\|$ really holds. Taking a first order approximation of the coupling term gives

$$\mathbf{a}(\tilde{\mathbf{x}}, t) - \mathbf{a}(\mathbf{x}, t) \approx \nabla \tilde{\mathbf{a}}(\tilde{\mathbf{x}}, t) \cdot (\tilde{\mathbf{x}}(t) - \mathbf{x}(t)) = \nabla \tilde{\mathbf{a}}(\tilde{\mathbf{x}}, t) \cdot \mathbf{e}(t) \ . \tag{50}$$

The gradient of the acceleration is dominated by the point-mass component of the gravity field. Thus, the characteristic value of the gradient $\nabla^* \mathbf{a}$ – the asterisk denotes a typical value – is

$$\nabla^* \mathbf{a} \approx \nabla^* \mathbf{g} \sim \frac{Gm_{Earth}}{\left(r^*\right)^3} \ . \tag{51}$$

Taking a characteristic radius $r^* \sim 20\,000$ km, gives $\nabla^* \mathbf{a} \sim 5 \cdot 10^{-8}$ s$^{-2}$. Assuming the target position error is met $\|\mathbf{e}\|^* \sim 1$ m and Eq. (50) gives $\|\mathbf{a}(\tilde{\mathbf{x}},t) - \mathbf{a}(\mathbf{x},t)\|^* \sim 5 \cdot 10^{-8}$ m/s$^2$. While this value is above our guesstimate for $\varepsilon_a$, it is still comparable (remember that we are interested only in the order of magnitude). Therefore, the derivation can be considered self-consistent and we may proceed forward assuming $\varepsilon_a = 10^{-8}$ m/s$^2$. This number must not be taken at face value, it is simply an initial guess. The suitability of this value shall be verified during the setup of the numerical model, to ensure that the target tolerances are met.

### 4.3 Tailoring the physical model

Once the threshold acceleration has been determined, it can be compared to the typical value of the different forces in action. Those that fall below the threshold can be removed from the model without compromising the accuracy of the calculations. For each source of perturbations, we shall determine the typical value at perigee and apogee.

In addition to the perturbation forces discussed in the previous section, it is worth checking the magnitude of the aerodynamic drag for completeness. At the perigee height – 1000 km – the atmospheric density depends strongly on solar activity, and is very difficult to predict. When the solar activity increases, the intermediate layers of the atmosphere expand, resulting in increased density at high altitude. According to the NRLMSISE-00 (Naval Research Laboratory Mass Spectrometer and Incoherent Scatter radar Exosphere) model [65] – the current standard for space research – the density at 1000 km varies roughly between $\rho = 10^{-15}$ kg/m$^3$ at times of low solar activity and $\rho = 10^{-14}$ kg/m$^3$ when the Sun is very active. The acceleration due to atmospheric drag is given by [66]

$$\mathbf{a}_{drag} = -\frac{1}{2}\rho \frac{A_{sat}}{m_{sat}} C_D \|\mathbf{v}_{sat}\| \mathbf{v}_{sat} \ , \tag{52}$$

where $C_D$ is the drag coefficient and $\mathbf{v}_{sat}$ is the speed of the satellite. The effect is obviously greatest at the perigee, where the maximum velocity – 9.65 km/s – and air density are encountered. Assuming $C_D \sim 1$ and high solar activity, the aerodynamic acceleration at perigee is $a_{drag} \sim 5 \cdot 10^{-16}$ m/s². This is eight orders of magnitude below $\varepsilon_a$, meaning that aerodynamic forces are irrelevant for the calculation.

|  | Typical acceleration (m/s²) | | |
| --- | --- | --- | --- |
| Perturbation | Perigee | Apogee | Above $\varepsilon_a = 10^{-8}$ m/s² |
| Lunar gravity | $1 \cdot 10^{-6}$ | $1 \cdot 10^{-5}$ | YES |
| Solar gravity | $6 \cdot 10^{-7}$ | $4 \cdot 10^{-6}$ | YES |
| Jovian gravity | $3 \cdot 10^{-12}$ | $2 \cdot 10^{-11}$ | NO |
| Solid Earth tide | $1 \cdot 10^{-7}$ | $1 \cdot 10^{-10}$ | YES |
| Ocean tide | $5 \cdot 10^{-9}$ | $1 \cdot 10^{-12}$ | NO |
| Sun radiation pressure | $5 \cdot 10^{-8}$ | $5 \cdot 10^{-8}$ | YES |
| Albedo radiation press. | $1 \cdot 10^{-8}$ | $2 \cdot 10^{-10}$ | MARGINALLY |
| Relativistic | $2 \cdot 10^{-8}$ | $2 \cdot 10^{-11}$ | MARGINALLY |
| Atmospheric drag | $5 \cdot 10^{-16}$ | $<< 5 \cdot 10^{-16}$ | NO |

**Table 1 - Characteristic accelerations due to different perturbation sources for a Molniya satellite**

Table 1 lists typical values of the perturbation sources. We see that the third-body effects due to Sun and Moon are clearly above the threshold, whereas the gravitational perturbation due to the Jovian system – computed for the January 1st 2021 epoch – is three orders of magnitude smaller than $\varepsilon_a$ and can be safely neglected.

The gravity from solid Earth tides is 10 times above the threshold at the perigee altitude, and should be taken into account. The overall effect over one orbit is likely limited, however, because the perturbation decays rapidly with height, becoming negligible at the apogee. Besides, the estimate in Table 1 is conservative, it corresponds to a Spring tide (Sun, Earth and Moon aligned). Most of the time the effect will be smaller. The contribution of ocean tides is always below $\varepsilon_a$, and is unlikely to have a significant effect at the target accuracy level.

The solar radiation pressure perturbation is almost an order of magnitude above $\varepsilon_a$ and must be included in the analysis. The albedo radiation perturbation barely reaches the cutoff value, so it is difficult to decide at this point if the effect is relevant. Section 5.2 contains a more detailed assessment. Please note that the estimated value in Table 1 is conservative because it assumes the Sun is directly at the zenith of the satellite, maximizing the effect of albedo.

Finally, the relativistic correction at the perigee height is also comparable to $\varepsilon_a$ – as always, we deal with approximations, the exact value is not important. Thus, whether it is relevant or not will be determined at a later stage.

In summary, a preliminary assessment suggests that the forces relevant for the problem are:

- Terrestrial gravity (field expansion degree adjusted for an accuracy of $10^{-8}$ m/s$^2$)
- Lunar gravity
- Solar gravity
- Solid Earth tides
- Solar radiation pressure
- Albedo radiation pressure
- Relativistic correction

The importance of each term shall be investigated in detail in the next section.

## 5  Numerical results

### 5.1  Integrator setup

Besides the physical model, the numerical parameters of the integrator must be adjusted in order to reach the desired accuracy. The embedded RK scheme is adaptive so, as far as precision is concerned, only the relative tolerance $\varepsilon_{rel}$ needs adjustment. The rest of the integrator parameters – like initial and maximum/minimum step sizes – have little effect on the accuracy (if set incorrectly, however, they might cause the propagation to terminate prematurely or increase the run time, but this is outside the scope of our discussion).

A priori, there is no well-defined relationship between the tolerance setting of the integrator and the error in the final state. The adaptive integrator adjusts the time increment to keep the error estimate – the difference between the high- and low-order approximations – below the specified tolerance for each time step. However, a typical simulation requires a large number of increments, so the individual step errors accumulate over time. The suitable tolerance setting must be determined empirically, studying the convergence of the final state. To this effect, we shall propagate the complete trajectory[2] – 60 revolutions, for which the target error is 180 m – with different tolerance levels and estimate the discretization error for each run (by comparison against the most accurate solution).

It is possible to obtain a rough estimate of the adequate relative tolerance setting by assuming that the accumulated absolute error per time step remains constant. Then, the error per step would be simply the final error divided by the number of steps. For a single revolution, numerical experiments show that the number of time steps used by the integrator is on the order of 50. This value depends weakly on the choice of tolerance, because for a method of orders 7/8 the time step size goes as the 8$^{th}$ root of the tolerance. Thus, one can compute a single revolution with a typical relative tolerance (say, $\varepsilon_{rel} = 10^{-8}$) and obtain a reasonable guess of the number of steps required, even if the tolerance used for the final calculation is different. Assuming 50 steps per revolution, 3000 time steps would be required for one month. Given that the final tolerable error is 180 m, hypothesizing that the error accumulates uniformly over time, the absolute error per step would be 6 cm. Thus, we can estimate the relative tolerance setting of the integrator as the ratio of error per step to the typical magnitude of the geocentric position vector (i.e., the relative position error accumulated during one step):

$$\varepsilon_{rel} \sim \frac{6\,\mathrm{cm}}{r^*} \sim 10^{-9} \ . \tag{53}$$

This value can be used to start the convergence study. Please note that it is just a rough orientation to commence the analysis.

To have an estimation of the exact solution – a requisite to compute errors – we shall propagate the trajectory including all the perturbative forces (even those that we expect to have negligible effects on the solution) and use a fixed degree *N*=100 for the synthesis of the gravity field. In Fig. 4 we see that the effective expansion degree of EGM2008 at the perigee altitude – 1000 km – is 71. Thus, additional harmonics would not improve the quality of the reference solution. An enhanced geopotential model (the soon-to-be released EGM2020, for example [67]) would be needed for better results.

---

[2] It is not necessary to use the complete trajectory for the convergence study or for analyzing the impact on the solution of the different perturbative forces (this is discussed in section 5.2). It would be possible to use a smaller number of revolutions for the preliminary calculations. For the sake of brevity, we used 60 orbits for all numerical experiments. This way the reference solutions from sections 5.1 and 5.2 can be used directly for section 5.3.

Next, we build a sequence of solutions with decreasing integrator tolerance, starting with the estimation (53), to determine at which level convergence is achieved. The estimation of the absolute position error of a solution corresponding to a tolerance setting $\varepsilon_{rel}$ is:

$$\tilde{\varepsilon}_x(T, \varepsilon_{rel}) = \left\| \mathbf{x}(T, \varepsilon_{rel}) - \mathbf{x}(T, \varepsilon_{rel}^{\min}) \right\|, \tag{54}$$

where $\varepsilon_{rel}^{\min}$ the minimum relative tolerance tested (corresponding to the most accurate solution). The orbit is not closed due to the perturbations, so the exact final position is unknown. Thus, the best approximation available is used instead.

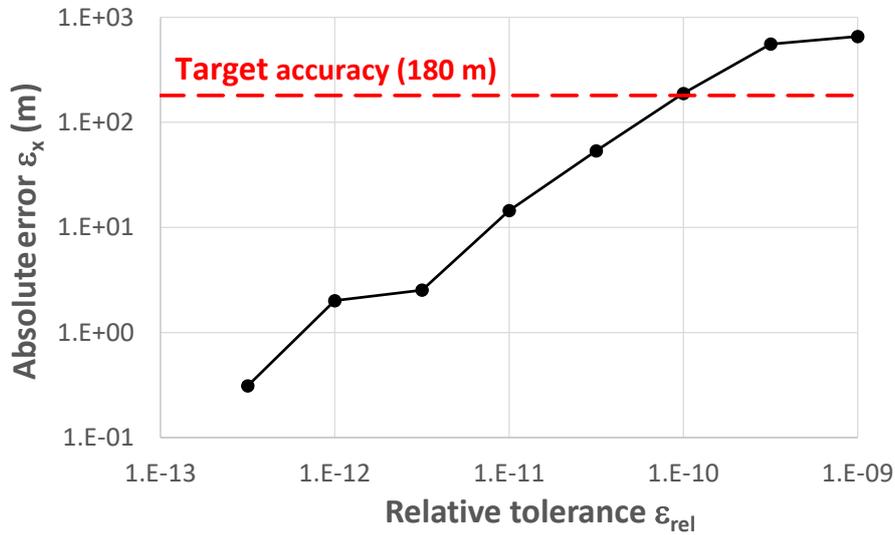

**Fig. 6 –Reference solution convergence (all perturbations, EGM expansion degree *N*=100)**

Fig. 6 displays the relationship between the integrator tolerance and positional error obtained for $\varepsilon_{rel}^{\min} = 10^{-13}$. Note that a suitable value of $\varepsilon_{rel}^{\min}$ cannot be chosen a priori, it has been determined empirically through examination of the results for different tolerances. Starting with the estimate (53), the relative tolerance is decreased in a geometric progression until the absolute error becomes small compared to 180 m. At that point, an adequate $\varepsilon_r^{\min}$ has been found. From Fig. 6 we can estimate that the error of the reference solution ($\varepsilon_{rel} = 10^{-13}$) is less than 1 m. This is small compared with the target accuracy and suitable to evaluate the true impact of each perturbative force (the focus of the next section).

## 5.2 Verification of the physical model

Once the discretization error due to the integrator has been reduced to a suitable level, the accuracy of the physical model can be assessed. To verify if the list of relevant perturbative forces sketched in section 4.3 is correct, we shall switch off one perturbation at a time to determine the accuracy loss with respect to the reference solution. Forces having an effect on the solution smaller than the absolute tolerance can be ignored safely.

Table 2 lists different combinations of physical model settings and the corresponding difference in final position compared to the reference configuration ($k=1$):

$$\delta_k = \|\mathbf{x}_k(T) - \mathbf{x}_1(T)\| .  \tag{55}$$

In all cases, the relative tolerance of the integrator is $10^{-13}$. Values of $\delta_k$ in bold characters indicate configurations not meeting the target accuracy (180 m).

| K | EGM setting | Sun 3rd body | Moon 3rd body | Jupiter 3rd body | Solar rad. | Albedo rad. | Earth tide | Ocean tide | Relativ. correc. | $\delta_k$ (m) |
|---|---|---|---|---|---|---|---|---|---|---|
| 1 | $N=100$ | ✓ | ✓ | ✓ | ✓ | ✓ | ✓ | ✓ | ✓ | - |
| 2 | $N=100$ | ✓ | ✓ | NO | ✓ | ✓ | ✓ | ✓ | ✓ | 0.12 |
| 3 | $N=100$ | ✓ | ✓ | ✓ | ✓ | ✓ | ✓ | NO | ✓ | 0.27 |
| 4 | $N=100$ | ✓ | ✓ | ✓ | ✓ | NO | ✓ | ✓ | ✓ | **380** |
| 5 | $N=100$ | ✓ | ✓ | ✓ | ✓ | ✓ | ✓ | ✓ | NO | 18 |
| 6 | $N=100$ | NO | ✓ | ✓ | ✓ | ✓ | ✓ | ✓ | ✓ | **35·10³** |
| 7 | $N=100$ | ✓ | NO | ✓ | ✓ | ✓ | ✓ | ✓ | ✓ | **12·10³** |
| 8 | $N=100$ | ✓ | ✓ | ✓ | NO | ✓ | ✓ | ✓ | ✓ | **1.3·10³** |
| 9 | $N=100$ | ✓ | ✓ | ✓ | ✓ | ✓ | NO | ✓ | ✓ | 49 |
| 10 | $10^{-9}$ m/s² | ✓ | ✓ | ✓ | ✓ | ✓ | ✓ | ✓ | ✓ | 1.1 |
| 11 | $10^{-8}$ m/s² | ✓ | ✓ | ✓ | ✓ | ✓ | ✓ | ✓ | ✓ | 71 |
| 12 | $10^{-7}$ m/s² | ✓ | ✓ | ✓ | ✓ | ✓ | ✓ | ✓ | ✓ | **380** |
| 13 | $10^{-8}$ m/s² | ✓ | ✓ | NO | ✓ | ✓ | NO | NO | NO | 140 |

**Table 2 – Change in final position of a Molniya orbit for different physical model configurations ($\varepsilon_{rel} = 10^{-13}$)**

Cases $k=2$ and $k=3$ confirm that effects of Jupiter's gravity and ocean tides are indeed negligible. They are smaller than 1 m, comparable to the accuracy of the numerical integration.

Cases 4 and 5 analyze the two perturbations that barely reached the acceleration threshold. We see that removing the relativistic correction is acceptable, as it only increases the error by 18 m. In contrast, neglecting the albedo radiation pressure causes an error of 380 m and is unacceptable.

Configurations 6 to 9 showcase the effect of those forces inducing accelerations above the threshold (in the case of solid Earth tides this happens only near the perigee). The gravity of the Sun and Moon causes displacements of tens of kilometers and must be retained, obviously. Likewise, ignoring solar radiation pressure yields an error above one kilometer. On the other hand, the solid Earth tide changes the final position by 50 m only, and can be removed from the model.

Cases 10 to 12 illustrate the effect of the dynamic EGM expansion degree. For a field truncation error $\varepsilon_{tru} = 10^{-8}$ m/s$^2$, as derived in section 4.2, the position error is below 70 m. For $\varepsilon_{tru} = 10^{-7}$ m/s$^2$ the error increases to 430 m, while for $\varepsilon_{tru} = 10^{-9}$ m/s$^2$ it drops to 1 m. Therefore, our initial guess of the required gravity synthesis accuracy was correct.

Inspecting the results for cases 2 to 11, we can finally determine the forces to include in the production model:

- Earth gravity to an accuracy of $10^{-8}$ m/s$^2$
- Solar gravity
- Lunar gravity
- Solar radiation pressure
- Albedo radiation pressure

Case 12 confirms that for this model the positional error is 140 m. Thus, it complies with our accuracy requirements.

The last parameter to determine is the relative tolerance of the integrator. From Fig. 6 we expect a setting of $10^{-11}$ would be adequate. It is worth noting that the convergence plot in Fig. 6 corresponds to the reference model (which includes all perturbation sources and uses a fixed expansion degree). In principle, it may not be applicable to the production configuration, which has a different physical model and dynamic field expansion. However, moving from a detailed model to a simpler one is not expected to impact negatively the performance of the integrator (on the contrary, the acceleration field of the reduced fidelity model is smoother, due to the lower degree of the spherical harmonics, and easier to propagate). Therefore, the relative tolerance guideline from Fig. 6 should be conservative when applied to the production model. To demonstrate this, we plotted the absolute error of the production model as a function of the integrator relative tolerance in Fig. 7. The error has been computed with respect to the reference solution (case k=1 in Table 2). It remains approximately constant for $\varepsilon_{rel} \leq 10^{-11}$. In this range the discretization error of the numerical integrator is small compared to the inaccuracy introduced by the simplified physics – 140 m, as shown by case 13 in Table 2 – and has little effect on the trajectory. For tolerances between $10^{-11}$ and $10^{-10}$ both sources of uncertainty are comparable and the curve becomes irregular. The dip in the curve is due to random error cancelations. For tolerances above $10^{-10}$ the discretization error of the integrator dominates, and the curve raises sharply. Henceforth, a relative tolerance of $10^{-11}$ shall be retained for the production model.

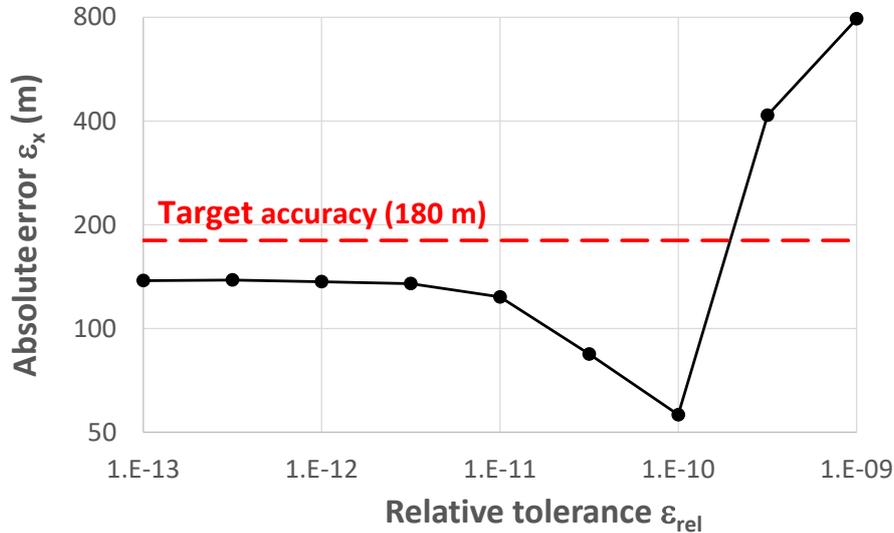

**Fig. 7 – Absolute error vs. integrator tolerance for production model**
**(Sun and Moon gravity, Solar and albedo radiation pressure, EGM accuracy $10^{-8}$ m/s2)**

## 5.3 Comparative performance and accuracy

We now discuss the accuracy and efficiency of the calculations. We present cost comparisons between the dynamic expansion degree technique and the conventional fixed-degree approach. The normalized computational cost (i.e., duration of the simulation) for each case is presented using the production configuration from section 5.2 as reference. To obtain a reliable estimate of the cost, the execution times are measured running the propagator for 1000 revolutions instead of 60. This averages out the effect of background activity on the host computer. We also showcase the ability of the dynamic degree choice to improve the solution accuracy with limited increases in solution time.

Table 3 presents a summary of the results. All the errors are measured with respect to the reference solution. Three configurations of the physical model are used:

- FULL: All perturbation forces included
- PROD: Production model, no tides, relativistic correction or Jupiter gravity
- ALT: Alternative model, perturbations included in PROD plus Earth tide and relativistic correction (an explanation for this choice is given below)

The first row of the table ($k$=1) is the reference solution used to estimate errors. Case $k$=14 (the numbering continues from Table 2 in order to have a unique identifier for each configuration) is the production configuration, which has unit normalized cost. Its absolute error is 125 m, within the tolerance limits. We see that the reference solution is 35 times more expensive.

In case 15 the production physical model is used together with a fixed-degree expansion $N$=100. The error target is met, as expected, but the cost is one order of magnitude higher than with the dynamic degree choice. The harmonic synthesis degree can be reduced to some extent without compromising the accuracy because, as we indicated before, it is above the effective expansion degree for the perigee altitude. To make a fair performance comparison, we reduced the expansion degree until the result became unreliable. For $N$=45 (case 16) the solution starts to fluctuate, indicating the truncation error becomes important. Further reductions of $N$ cause unpredictable increases and decreases of the error, due to random cancellations. Even though case 16 has been carefully tuned for maximum performance, it remains two and a half times more expensive than the dynamic expansion ($k$=14). Clearly, a fixed expansion degree must cannot deliver optimal performance when the satellite experiences large variations of altitude.

To further illustrate the benefits of the dynamic expansion technique, we computed a higher accuracy solution ($k$=17) by including all the perturbation sources and lowering the EGM truncation degree to $10^{-9}$ m/s$^2$. A reduction of the integrator tolerance to $10^{-12}$ was required to realize the benefits of the improved physical model. We see that the errors drops by two orders of magnitude while the cost increases to 2.8 only. In contrast, the same computation using fixed $N$=100 ($k$=18) has a cost index of 15. Again, for the sake of fairness, we reduced the expansion degree to the point where the accuracy begins to degrade ($N$=60, $k$=19). The accuracy (about 1 m) is preserved, but the cost index remain elevated at 6.6. For illustrative purposes, we further reduced the expansion degree until the cost became comparable to the configuration $k$=17. This is case $k$=20 (expansion degree N=20) which incurs an error of 50 m, totally negating to benefits of the improved physical model.

A benefit of the sensitivity study from section 4.3 (Table 2) is that we can easily improve the performance of the solution by removing physical effects not relevant at the current accuracy level (1 m for the high-accuracy case). The table shows that removing Jupiter's gravity and ocean tides would introduce an error on the order of 10 cm, which is acceptable. This is important from the performance standpoint, because the computation of the ocean tide is expansive due to the large number of constituents in FES2004 [55]. We ran a computation ($k$=21) which ignores these two perturbations (ALT physics model). Using variable expansion degree we achieve the same accuracy as case 17, but with a cost index of 1.9. Of course, the same strategy can be applied to fixed expansion degree ($k$=22) but the cost index only drops to 5.4 (2.8 times higher than the equivalent dynamic expansion case, $k$=21).

| $k$ | Comment | EGM setting | $\varepsilon_{rel}$ | Physics model | $\delta_k$ (m) | Cost index |
|---|---|---|---|---|---|---|
| 1 | Reference | $N$=100 | $10^{-13}$ | FULL | - | 19 |
| 14 | Production | $\varepsilon_{tru}$ =$10^{-8}$ m/s$^2$ | $10^{-11}$ | PROD | 125 | 1 |
| 15 | PROD physics, $N$=100 | $N$=100 | $10^{-11}$ | PROD | 54 | 9.6 |
| 16 | PROD physics, $N$=45 | $N$=45 | $10^{-11}$ | PROD | 49 | 2.4 |
| 17 | High accuracy, Dyn. degree | $\varepsilon_{tru}$ =$10^{-9}$ m/s$^2$ | $10^{-12}$ | FULL | 1.8 | 3.2 |
| 18 | FULL physics, $N$=100 | $N$=100 | $10^{-12}$ | FULL | 0.89 | 15 |
| 19 | FULL physics, $N$=60 | $N$=60 | $10^{-12}$ | FULL | 1.4 | 6.6 |
| 20 | FULL physics, $N$=30 | $N$=30 | $10^{-12}$ | FULL | 50 | 3.3 |
| 21 | ALT physics, Dyn. degree | $\varepsilon_{tru}$ =$10^{-9}$ m/s$^2$ | $10^{-12}$ | ALT | 1.8 | 1.9 |
| 22 | ALT physics, $N$=60 | $N$=60 | $10^{-12}$ | ALT | 1.6 | 5.4 |
| 23 | Low fidelity, Dyn. degree | $\varepsilon_{tru}$ =$10^{-7}$ m/s$^2$ | $10^{-11}$ | PROD | 160 | 0.83 |
| 24 | Low fidelity, $N$=20 | $N$=20 | $10^{-11}$ | PROD | **240** | 0.91 |

**Table 3 – Comparison of accuracy and computational cost for different model setups of a Molniya orbit**

Finally, we tried a reduced fidelity computation increasing the EGM truncation error to $10^{-9}$ m/s$^2$ (case 23). The error remains almost unchanged − and within requirements − while achieving a cost reduction of 20%. On the other hand, if we try to build a fixed-degree solution having comparable cost ($k$=24) the error jumps to 240 m, outside tolerance. Once again, the fixed expansion degree does not strike a good balance between accuracy and cost.

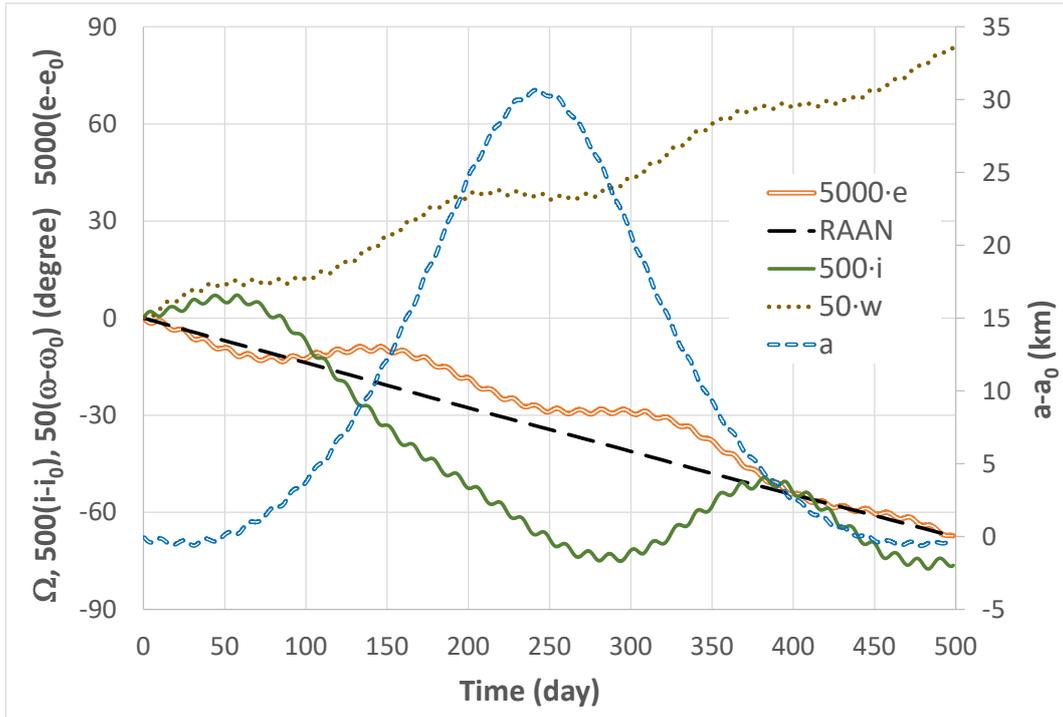

**Fig. 7 – Change in orbital elements over 500 days (reference solution)**

For informative purposes, we included the evolution of the orbital elements for the reference solution ($k=1$) in Fig. 7. Note that the changes in inclination ($i$), eccentricity ($e$) and argument of the perigee ($\omega$) have been upscaled to make them apparent, because they are very small compared to the variation of RAAN. The latter changes by 70⁰ over 500 days due to Earth's polar flattening. In contrast, over the same time span, $\omega$ increases by less than 2⁰ due to the choice of critical inclination, which cancels the effect of Earth's oblateness. The changes of inclination and eccentricity, mostly due to lunisolar perturbations, are very small (0.15⁰ and 0.014⁰, respectively). Finally, the variation of the semimajor axis, which is quasi-periodic, has a small relative amplitude (30 km). However, it is important for satellite phasing because it affects the orbital period.

## 6 CONCLUSIONS

We presented a simple and effective method for optimizing the precision and efficiency of an orbit propagator. Starting with the accuracy requirement of the application (which is a function of the mission objectives), the allowable acceleration error is determined. The physical model is then adapted to match this uncertainty, removing as much complexity from it as possible. Together with a proper setting of the integrator tolerance, this yields a solution that respects the accuracy constraints of the problem while minimizing the computational cost. The preparation of the model only needs back-of-the-envelope calculations and some propagation tests. These can be completed in a short time, optimizing the performance of the production model (especially important for sensitivity analysis and optimization, where large batches of cases need to be computed) and giving the analyst confidence in the results.

A vital part of the strategy involves an analysis of the uncertainty of the gravity field model. From the statistical uncertainty of the Stokes coefficients, the typical error of the gravitational acceleration is derived as a function of height. This defines the accuracy limit that can be achieved from a particular geopotential model. Using this result, an analyst can quickly determine if a particular EGM is suitable for his application.

A study of the truncation error of the field as a function of the expansion degree shows that, for a given precision, the expansion degree degreases rapidly with altitude ($N_{req} \propto h^{-1}$, approximately). It is clear that, for orbits where height is not constant, a fixed expansion degree is a suboptimal choice. Our study for the Molniya orbit shows that by using a dynamic degree selection, the accuracy of the solution can be maintained while reducing the computational cost by a factor of 3 or more. Moreover, when the dynamic expansion is used, accuracy can be improved with a modest change in solution cost. We managed to increase the accuracy of the baseline solution by two orders of magnitude with just a 90% rise in computational cost.

Even in the case of circular orbits – where the dynamic expansion offers no efficiency advantage – our results are important, because they remove the guesswork from the selection of the truncation degree. This is a marked improvement with respect to generic recommendations, that do not consider the specific requirements of each application. For example, the IERS Conventions 2010 [54] gives the following guidelines for EGM2008 truncation levels:

| Orbit height (km) | Satellite | Truncation level |
|---|---|---|
| 953 | Starlette | 90 |
| 5892 | LAGEOS | 20 |
| 20222 | GPS | 12 |

**Table 4 – EGM2008 suggested truncation levels from IERS Conventions 2010 [54]**

The suggested truncation levels in Table 4 agree very well with the effective expansion degree $N_{eff}(h)$ from Fig. 4. This is unsurprising because the recommendation is meant for geodesy work, where maximum accuracy is sought. For less stringent applications, using the IERS recommendation would be computationally inefficient. Our methodology offers additional flexibility, allowing the analyst to adjust the gravity expansion degree for any specific application.

## ACKNOWLEDGEMENTS

This work has been funded by Khalifa University of Science and Technology's internal grants FSU-2018-07 and CIRA-2018-85. R. Flores also acknowledges financial support from the Spanish Ministry of Economy and Competitiveness, through the "Severo Ochoa Programme for Centres of Excellence in R&D" (CEX2018-000797-S).

## REFERENCES

[1] A. Rossi (2004), Population models of space debris, Proceedings of the International Astronomical Union, (IAUC197), 427-438 (Z. Knezevic and A. Milani, eds.) DOI: 10.1017/S1743921304008956

[2] D. Vallado, Fundamentals of Astrodynamics and Applications, Springer, New York, 2007

[3] O. Montenbruck, E. Gill, Satellite Orbits: Models, Methods and Applications, Springer, Berlin, Germany, 2000

[4] J.M.A. Danby, Fundamentals of Celestial Mechanics, Willmann-Bell, Richmond, VA, 1992

[5] J.F. Herman, B.A. Jones, G.H. Born, J.S. Parker (2013), A comparison of implicit integration methods for astrodynamics, AAS/AIAA Astrodynamics Specialist Conference, Hilton Head, SC, Paper AAS 13-905

[6] B.A. Jones, R.L. Anderson (2012), A survey of symplectic and collocation integration methods for orbit propagation, AAS/AIAA Space Flight Mechanics Meeting, Charleston, SC, Paper AAS 12-214

[7] O. Montenbruck, Numerical integration methods for orbital motion, Celestial Mechanics and Dynamical Astronomy 53 (1992), 59-69, DOI: 10.1007/BF00049361

[8] K. Fox, Numerical integration of the equations of motion of celestial mechanics, Celestial Mechanics 33 (1984), 127–142, DOI: 10.1007/BF01234151

[9] A. Deprit, The main problem in the Theory of Artificial Satellites to Order Four, Journal of Guidance and Control 4(2) (1981), 201-206, DOI: 10.2514/3.56072

[10] W.M. Kaula, Theory of Satellite Geodesy, Blaisdell Publishing Co., Waltham, MA, 1966

[11] Y. Kozai, Second-Order Solution of Artificial Satellite Theory without Drag, Astronomical Journal 67, 446-461 (1962), DOI: 10.1086/108753

[12] D. Brouwer, Solutions of the Problem of Artificial Satellite Theory without Drag, Astronomical Journal 64 (1959), 378-397, DOI: 10.1086/107958

[13] F.R. Hoots, R.G. France, An Analytic Satellite Theory using Gravity and a Dynamic Atmosphere, Celestial Mechanics 40 (1987), 1-18, DOI: 10.1007/BF01232321

[14] J. Liu, Satellite Motion about an Oblate Earth, AIAA Journal 12 (1974), 1511-1516, DOI: 10.2514/3.49537

[15] P.J. Cefola, A.C. Long, G. Halloway (1974), The Long-Term Prediction of Artificial Satellite Orbits, AIAA Aerospace Sciences Meeting, Washington DC, Paper AIAA-74-170

[16] N. Z. Miura, Comparison and Design of Simplified General Perturbation Models, California Polytechnic State University, San Luis Obispo, 2009

[17] H. Fraysse, V. Morand, C. Le Fevre et al. (2011), Long-Term Orbit Propagation Techniques Developed in the Frame of the French Space Act, 22$^{nd}$ International Symposium on Space Flight Dynamics, São José dos Campos, Brazil

[18] M. Lara, J.F. San-Juan, D. Hautesserres, HEOSAT: a mean elements orbit propagator program for highly elliptical orbits, CEAS Space Journal 10 (2018), 3–23, DOI: 10.1007/s12567-017-0152-x

[19] J.M. Aristoff, J.T. Horwood, A.B. Poore, Orbit and uncertainty propagation: a comparison of Gauss–Legendre-, Dormand–Prince-, and Chebyshev–Picard-based approaches, Celestial Mechanics and Dynamical Astronomy 118 (2014), 13-28, DOI: 10.1007/s10569-013-9522-7

[20] J.M. Aristoff, A.B. Poore (2012), Implicit Runge-Kutta methods for orbit propagation, AIAA/AAS Astrodynamics Specialist Conference, Minneapolis, MN, Paper AIAA 2012-4880

[21] S. Blanes, F. Casas, A. Farrés et al., New families of symplectic splitting methods for numerical integration in dynamical astronomy, Applied Numerical Mathematics 68 (2013), 58-72, DOI: 10.1016/j.apnum.2013.01.003

[22] A. Farrés, J. Laskar, S. Blanes et al., High precision symplectic integrators for the Solar System, Celestial Mechanics and Dynamical Astronomy 116 (2013), 141–174, DOI: 10.1007/s10569-013-9479-6

[23] S. Mikkola, Efficient symplectic integration of satellite orbits, Celestial Mechanics and Dynamical Astronomy 74(4) (1999), 275-285, DOI: 10.1023/A:1008398121638

[24] B.A. Jones, R.L. Anderson (2012), A survey of symplectic and collocation integration methods for orbit propagation, AAS/AIAA Space Flight Mechanics Meeting, Charleston, SC, Paper AAS 12-214

[25] G. Baù, A. Hunh, H. Urrutxua, C. Bombardelli, J. Peláez (2011), DROMO: a new regularized orbital propagator, International Symposium on Orbit Propagation and Determination, IMCCE, Lille, France

[26] D. Amato, C. Bombardelli, G. Baù, et al., Non-averaged regularized formulations as an alternative to semi-analytical orbit propagation methods, Celestial Mechanics and Dynamical Astronomy 131 (2019), 21, DOI: 10.1007/s10569-019-9897-1

[27] G. Baù, C. Bombardelli, J. Peláez and E. Lorenzini, Non-singular orbital elements for special perturbations in the two-body problem, Monthly Notices of the Royal Astronomical Society 454(3) (2015), 2890–2908, DOI: 10.1093/mnras/stv2106

[28] G. Baù, C. Bombardelli, J. Peláez (2013), Accurate and Fast Orbit Propagation with a New Complete Set of Elements, AAS/AIAA Space Flight Mechanics Meeting, Kauai, Hawaii, Paper AAS 13-491

[29] B.K. Bradley, B. Jones, G. Beylkin, P. Axelrad, A new numerical integration technique in astrodynamics, Advances in the Astronautical Sciences, 143 (2012), 1709-1728

[30] M.M. Berry, L.M. Healy, Implementation of Gauss-Jackson integration for orbit propagation, Journal of the Astronautical Sciences 52(3) (2004), 331-357

[31] M. Lara (2018), Exploring Sensitivity of Orbital Dynamics with Respect to Model Truncation: The Frozen Orbits Approach, Stardust Final Conference, Astrophysics and Space Science Proceedings 52, DOI: 10.1007/978-3-319-69956-1_4

[32] K. Sośnica, D. Thaller, A. Jäggi, R. Dach, G. Beutler, Sensitivity of Lageos orbits to global gravity field models, Artificial Satellites 47(2) (2012), DOI: 10.2478/v10018-012-0013-y

[33] R. D. Brouwer, G.M. Clemence, Methods of celestial mechanics, Academic Press, New York, USA, 1961

[34] W. Heiskanen, H. Moritz, Physical Geodesy, Freeman, San Francisco, CA, 1967

[35] F.G. Lemoine, S.C. Kenyon, J.K. Factor et al., The Development of the Joint NASA GSFC and the National Imagery and Mapping Agency (NIMA) Geopotential Model EGM96, Technical Publication TP-1998-206861, NASA Goddard Space Flight Center, 1998

[36] N.K. Pavlis, S.A. Holmes, S.C. Kenyon, et al., The development and evaluation of the Earth gravitational model (EGM2008), J. Geophys. Res. 117(B04406) (2008), DOI: 10.1029/2011JB008916

[37] E. Fantino, R. Flores, M. Di Carlo, et al., Geosynchronous inclined orbits for high-latitude communications, Acta Astronautica 140 (2017), 570-582, DOI: 10.1016/j.actaastro.2017.09.014

[38] E. Fantino, R. Flores, A. Adheem (2019), Accurate and Efficient Propagation of Satellite Orbits in the Terrestrial Gravity Field, 70[th] International Astronautical Congress, Washington D.C., USA, Paper IAC-19-C1.3.8

[39] Y. Kolyuka, N. Ivanov, T. Afanasieva, T. Gridchina (2009), Examination of the lifetime, evolution and re-entry features for the "Molniya" type orbits, 21st International Symposium on Space Flight Dynamics, Toulouse, France

[40] E. Fehlberg, Classical Fifth-, Sixth-, Seventh-, and Eight-Order Runge-Kutta Formulas with Stepsize Control, Technical report TR R-287, NASA Johnson Space Center, 1968

[41] S.E. Urban, P.K. Seidelmann, Explanatory Supplement to the Astronomical Almanac, 3rd Ed., University Science Books, Herndon, VA, 2012, ISBN: 978-1891389856

[42] R. H. Battin, An Introduction to the Mathematics and Methods of Astodynamics, Revised Ed., AIAA Education Series, Reston-VA, USA, 1999, ISBN 978-1563473425

[43] NASA JPL HORIZONS Web-Interface: https://ssd.jpl.nasa.gov/horizons.cgi (accessed December 10th 2020)

[44] J.H. Lieske, T. Lederle, W. Fricke, B. Morando, Expressions for the precession quantities based upon the IAU (1976) system of astronomical constants, Astronomy and Astrophysics 58(1-2) (1977), 1-16

[45] J.L. Hilton, N. Capitaine, J. Chapront et al., Report of the International Astronomical Union Division I Working Group on Precession and the Ecliptic, Celestial Mechanics and Dynamical Astronomy 94 (2006), 351–367, DOI: 10.1007/s10569-006-0001-2

[46] L.V. Morrison, F.R. Stephenson, Historical values of the Earth's clock error ΔT and the calculation of eclipses, Journal for the History of Astronomy, 94(120) (2004), 327-336, DOI: 10.1177/002182860503600307

[47] N. Capitaine, P.T. Wallace, J. Chapront, Improvement of the IAU 2000 precession model, Astronomy and Astrophysics 432(1) (2005), 355-367, DOI: 10.1051/0004-6361:20041908

[48] S. A. Holmes, W.E. Featherstone, A unified approach to the Clenshaw summation and the recursive computation of very high degree and order normalised associated Legendre functions, Journal of Geodesy 76 (2002), 279–299, DOI: 10.1007/s00190-002-0216-2

[49] Intel® 64 and IA-32 Architectures Software Developer's Manual Volume 1: Basic Architecture, Order Number: 253665-060US, Intel corporation, USA, 2016. Available online at: https://www.intel.com/content/dam/www/public/us/en/documents/manuals/64-ia-32-architectures-software-developer-vol-1-manual.pdf (accessed May 15th 2020)

[50] S. Pines, Uniform representation of the gravitational potential and its derivatives, AIAA J. 11 (1973), 1508–1511, DOI: 10.2514/3.50619

[51] G. Balmino, J.-P. Barriot, N. Vales, Non-singular formulation of the gravity vector and gravity gradient tensor in spherical harmonics, Manuscripta Geod. 15(1) (1990), 11–16, DOI: 10.1023/A:1008361202235

[52] G. Balmino, J.-P. Barriot, B. Koop, et al., Simulation of gravity gradients: a comparison study, Bull. Geodesique 65(4) (1991), 218–229, DOI: 10.1007/BF00807265

[53] E. Fantino, S. Casotto, Methods of harmonic synthesis for global geopotential models and their first-, second- and third-order gradients, J. Geodesy 83(4) (2009), 595–619, DOI: 10.1007/s00190-008-0275-0

[54] G. Petit, B. Luzum (eds.), IERS Conventions (2010). (IERS Technical Note ; 36) Frankfurt am Main: Verlag des Bundesamts für Kartographie und Geodäsie, 2010, ISBN: 3-89888-989-6

[55] F. Lyard, F. Lefevre, T. Letellier, O. Francis, Modelling the global ocean tides: modern insights from FES2004, Ocean Dynamics 56 (2006), 394–415, DOI: 10.1007/s10236-006-0086-x.

[56] E.W. Weisstein, "Lune." From MathWorld - A Wolfram Web Resource. https://mathworld.wolfram.com/Lune.html (accessed December 10$^{th}$ 2020)

[57] N. Borderies, Py. Longaretti, A new treatment of the albedo radiation pressure in the case of a uniform albedo and of a spherical satellite. Celestial Mechanics and Dynamical Astronomy 49 (1990), 69–98, DOI: 10.1007/BF00048582L

[58] G. Bury, K. Sośnica, R. Zajdel et al., Toward the 1-cm Galileo orbits: challenges in modeling of perturbing forces. J. Geodesy 94(16) (2020), DOI: 10.1007/s00190-020-01342-2

[59] L. Prange, A. Villiger, D. Sidorov, et al., Overview of CODE's MGEX solution with the focus on Galileo, Advances in Space Research 66(12) (2020), 2786–2798, DOI: 10.1016/j.asr.2020.04.038

[60] R. Fitzpatrick, An Introduction to Celestial Mechanics, Cambridge University Press, Cambridge, UK, 2012

[61] R. Shako et al., EIGEN-6C: A High-Resolution Global Gravity Combination Model Including GOCE Data. In: F. Flechtner, N. Sneeuw, W.D. Schuh, Observation of the System Earth from Space - CHAMP, GRACE, GOCE and future missions. Advanced Technologies in Earth Sciences. Springer, Berlin, 2014, DOI: 10.1007/978-3-642-32135-1_20

[62] P. Billingsley, Probability & Measure, 3$^{rd}$ Ed., Wiley, New York, USA, 1995, ISBN 0-471-007-02

[63] C.F. Gauss, Exposition d'une Nouvelle Méthode de Calculer les Perturbations Planétaires avec L'application au Calcul Numérique des Perturbations du Mouvement de Pallas, Werke. Hrsg. Gesellschaft der Wissenschaften zu Göttingen, K. von der Göttingen, 7 (1870), 439–472

[64] J.F. Encke, Über die allgemeinen Störungen der Planeten, Berliner Astronomisches Jahrbuch für 1857 (1854), 319–397

[65] J.M. Picone, A.E. Hedin, D.P. Drob, A.C. Aikin, NRLMSISE-00 empirical model of the atmosphere: Statistical comparisons and scientific issues, Journal of Geophysical Research, 107(A12) (2002), SIA15-1–SIA15-16, DOI: 10.1029/2002JA009430

[66] A. Tewari, Atmospheric and Space Flight Dynamics, Birkhäuser, Boston, 2007

[67] D. Barnes, S. Holmes, J. Factor el al. (2017), Earth Gravitational Model 2020, Proceedings from the 19th EGU General Assembly, Vienna, Austria